\documentclass[conference]{IEEEtran}
\makeatletter
\def\ps@headings{%
\def\@oddhead{\mbox{}\scriptsize\rightmark \hfil \thepage}%
\def\@evenhead{\scriptsize\thepage \hfil \leftmark\mbox{}}%
\def\@oddfoot{}%
\def\@evenfoot{}}
\makeatother 

\usepackage{graphicx, url}
\usepackage{subfigure}
\usepackage{color}
\usepackage{epsfig}
\usepackage{amssymb}
\usepackage{latexsym}
\usepackage{amsmath}

\newcommand {\C} {{\rm I\kern-5.5pt C}}

\newcommand{\bP}[1]{{\mathbb{P}}\left[{#1}\right]}
\newcommand{\bE}[1]{{\mathbb{E}}\left[{#1}\right]}

\newcommand{\1}[1]{{\bf 1}\left[#1\right]}       

\newcommand{\fsquare}{\vrule height6pt width7pt depth1pt}   
\newcommand{\myproof}{{\hfill \\ \bf Proof. \ }}           
\newcommand{\myendpf}{\hfill\fsquare \\[0.1in]}             

\newtheorem{theorem}{Theorem}[section]

\newtheorem{lemma}[theorem]{Lemma}
\newtheorem{proposition}[theorem]{Proposition}

\begin{document}

\title{Performance of the Eschenauer-Gligor key distribution scheme
  under an ON/OFF channel
      }

\author{
\authorblockN{Osman Ya\u{g}an}
\authorblockA{Department of Electrical and Computer Engineering\\
              and the Institute for Systems Research\\
              University of Maryland, College Park\\
              College Park, Maryland 20742\\
              osmanyagan@gmail.com}
}

\maketitle

\begin{abstract}
\normalsize We investigate the secure connectivity of wireless
sensor networks under the random key distribution scheme of
Eschenauer and Gligor. Unlike recent work which was carried out
under the assumption of {\em full visibility}, here we assume a
(simplified) communication model where unreliable wireless links
are represented as on/off channels. We present conditions on how
to scale the model parameters so that the network i) has no secure
node which is isolated and ii) is securely connected, both with
high probability when the number of sensor nodes becomes large.
The results are given in the form of full {\em zero-one laws}, and
constitute the first {\em complete} analysis of the EG scheme
under {\em non}-full visibility. Through simulations these
zero-one laws are shown to be valid also under a more realistic
communication model, i.e., the disk model. The relations to the
Gupta and Kumar's conjecture on the connectivity of geometric
random graphs with randomly deleted edges are also discussed.
\end{abstract}


\section{Introduction}
\label{sec:Introduction}

\subsection{Wireless sensor networks and security}

Wireless sensor networks (WSNs) are distributed collections of
sensors that are envisioned \cite{Akyildiz} to be used in a wide
range of application areas including healthcare (e.g.  patient
monitoring), military operations (e.g., battlefield surveillance)
and homes (e.g., home automation and monitoring). These WSNs will
often be deployed in hostile environments where communications can
be monitored, and nodes are subject to capture and surreptitious
use by an adversary. Under such circumstances, cryptographic
protection will be needed to ensure secure communications, and to
support functions such as sensor-capture detection, key revocation
and sensor disabling.

Unfortunately, many security schemes developed for general network
environments do not take into account the unique features of WSNs:
Public key cryptography is not computationally feasible because of
the severe limitations imposed on the physical memory and power
consumption of the individual sensors. Traditional key exchange
and distribution protocols are also not useful as they are based
on trusting third parties while the topologies of large-scale WSNs
are unknown prior to deployment. We refer the reader to the papers
\cite{CamtepeYener,EschenauerGligor,PerrigStankovicWagner,SunHe}
for more detailed discussions on the security challenges in WSN
settings.

{\em Random} key predistribution schemes were recently introduced
to address some of these difficulties. The idea of randomly
assigning secure keys to sensor nodes prior to network deployment
was first introduced by Eschenauer and Gligor
\cite{EschenauerGligor}. According to their scheme, here after
referred to as the EG scheme, each sensor is independently
assigned $K$ distinct cryptographic keys which are selected
uniformly at random from a pool of $P$ keys. These $K$ keys
constitute the key ring of the node and are inserted into its
memory before the network deployment. Two sensor nodes can then
establish a secure link between them if they are within
transmission range of each other and if their key rings have at
least one key in common; see \cite{EschenauerGligor} for
implementation details.

Since then, many competing alternatives to the EG scheme have been
proposed; see \cite{CamtepeYener} for a detailed survey of various
key distribution schemes for WSNs. With a number of schemes
available, a basic question arises as to how they compare with
each other. Answering this question requires a good understanding
of the properties and performance of the schemes under
consideration, and there are a number of ways to achieve this. The
approach we use here considers random graph models naturally
induced by a given scheme, and then develops the scaling laws
corresponding to desirable network properties, e.g., absence of
secure nodes which are isolated, secure connectivity, etc. This is
done with the aim of deriving guidelines to {\em dimension} the
scheme, namely adjust its parameters so that these properties
occur with high probability as the number of nodes becomes large.
Here, we focus on the connectivity properties since secure
connectivity is one of the basic properties required for a
successful operation of the WSN.

\subsection{Relevant work}

To date, much efforts along the above lines have been carried out
under the assumption of {\em full visibility} according to which
sensor nodes are all within communication range of each other.
Under this assumption, the EG scheme gives rise to a class of
random graphs known as random key graphs; relevant results are
available in the references \cite{BlackburnGerke,
DiPietroManciniMeiPanconesiRadhakrishnan2008, EschenauerGligor,
Rybarczyk2009, YaganMakowskiConnectivity}. The q-composite scheme
\cite{ChanPerrigSong}, a simple variation of the EG scheme, was
investigated by Bloznelis et al. \cite{BloznelisJaworskiRybarczyk}
through an appropriate extension of the random key graph model.
Recently, Ya\u{g}an and Makowski have analyzed various random
graphs induced by the random pairwise key predistribution scheme
of Chan et al. \cite{ChanPerrigSong}; see
\cite{YaganMakowskiICC2011}.

To be sure, the full visibility assumption does away with the
wireless nature of the communication medium supporting WSNs. In
fact, a common criticism of the above line of work is that by
disregarding the unreliability of the wireless links, the
resulting dimensioning guidelines are likely to be too {\em
optimistic}: In practice nodes will have fewer neighbors since
some of the communication links may be impaired. As a result, the
desired connectivity properties may not be achieved if
dimensioning is done according to results derived under full
visibility.

With this in mind, there has been a number of efforts to
incorporate a wireless communication model to the existing full
visibility models of the key distribution schemes. Among them, the
most popular one is the so called disk model \cite{GuptaKumar}:
Assuming that the sensors are distributed over a bounded region
$\mathcal{D}$ of a euclidian plane, two nodes are assumed to have
a direct communication link in between as long as they are within
transmission range of each other. In other words, with $\rho >0$
denoting the transmission range, nodes $i$ and $j$ located at
$\boldsymbol{x_i}$ and $\boldsymbol{x_j}$ are able to communicate
if
\begin{equation}
\parallel \boldsymbol{x_i} -\boldsymbol{x_j} \parallel<\rho.
\label{eq:cond_wireless_range}
\end{equation}
When the node locations are independently and uniformly
distributed over the region $\mathcal{D}$, the graph induced under
the condition (\ref{eq:cond_wireless_range}) is known as a random
geometric graph \cite{GuptaKumar,PenroseBook} for which the most
well-known result is the following zero-one law for connectivity
\cite{GuptaKumar}: If $\mathcal{D}$ is a disk of unit area, $\rho$
is scaled with the number of nodes $n$, and it holds that
\[
\pi \rho_n^2 =\frac{\log n + w_n}{n},
\]
then the probability that the resulting geometric random graph is
connected tends to $1$ (resp. $0$) as $n$ gets large if $\lim_{n
\to \infty} w_n =\infty$ (resp. $\lim_{n \to \infty} w_n = -
\infty$). It was also conjectured by Gupta and Kumar
\cite{GuptaKumar} that if each edge of the geometric random graph
was to be deleted with probability $1-\alpha$ independently from
all the other edges, then the zero-one law for connectivity would
take the following form: If $\alpha$ and $\rho$ are scaled with
$n$ and it holds that
\begin{equation}
\pi \rho_n^2 \alpha_n =\frac{\log n + w_n}{n},
\label{eq:conjecture_gupta_kumar}
\end{equation}
then the resulting random graph, which is an {\em intersection} of
the random geometric graph and the Erd\H{o}s-R\'enyi graph
\cite{Bollobas}), is connected with probability tending to $1$
(resp. $0$) if $\lim_{n \to \infty} w_n =\infty$ (resp. $\lim_{n
\to \infty} w_n = - \infty$).

Inspired by these, the studies on the secure connectivity of WSNs
have focused \cite{KrzywdziRybarczyk,YiWanLinHuang} on
establishing an appropriate analog of the conjecture
(\ref{eq:conjecture_gupta_kumar}). After all, incorporating the
disk model to the EG scheme corresponds to studying a random graph
formed by {\em intersecting} the geometric random graph with the
random key graph. As a result, one can conjecture that, with
$\beta$ denoting the probability that two nodes have at least one
common key in the EG scheme, and
\begin{equation}
\pi \rho_n^2 \beta_n =\frac{\log n + w_n}{n},
\label{eq:conjecture_OY}
\end{equation}
we have connectivity with probability tending to $1$ (resp. $0$)
as $n$ gets large if $\lim_{n \to \infty} w_n =\infty$ (resp.
$\lim_{n \to \infty} w_n = - \infty$).

To date, both of the conjectures (\ref{eq:conjecture_gupta_kumar})
and (\ref{eq:conjecture_OY}) remain to be open. In fact, despite
several attempts, even the less stronger forms of the conjectures
have not been established yet. Namely, the conjectures
(\ref{eq:conjecture_gupta_kumar}) and (\ref{eq:conjecture_OY})
imply that with
\begin{eqnarray}
\label{eq:conjecture_gupta_kumar_weak} &&\pi \rho_n^2 \alpha_n
=c\frac{\log n}{n}
\\
&& \pi \rho_n^2 \beta_n =c\frac{\log n}{n},
\label{eq:conjecture_OY_weak}
\end{eqnarray}
respectively, the resulting graphs are connected with probability
tending to $1$ (resp. $0$) if $c>1$ (resp. $c<1$).

 For instance,
Di Pietro et al.
\cite{DiPietroMeiManciniPanconesiRadhakrishnan2004} have shown
that under the scaling (\ref{eq:conjecture_OY_weak}), the one law
$\lim_{n \to \infty} \bP{\textrm{Corresponding random graph is
connected}} = 1$ follows if $\rho_n>0$ and $c>20\pi$. Very
recently, Krzywdzi\'nski and Rybarczyk \cite{KrzywdziRybarczyk}
have improved this results and established the one-law under
(\ref{eq:conjecture_OY_weak}) for $c>8$ without any constraint on
$\rho_n$. In \cite{KrzywdziRybarczyk}, the authors have also
established the one-law under
(\ref{eq:conjecture_gupta_kumar_weak}) for $c>8$ marking the first
connectivity result for the random geometric graph with random
edge deletion. Another notable work is due by Yi et al.
\cite{YiWanLinHuang}, where the authors have established the
strong forms (\ref{eq:conjecture_gupta_kumar}) and
(\ref{eq:conjecture_OY}) of the conjectures but {\em only} for the
property of absence of isolated nodes; there, it was also assumed
that $\lim_{n \to \infty} \alpha_n \log n = \infty$ and $\lim_{n
\to \infty} \beta_n \log n = \infty$. Clearly, absence of isolated
nodes is a necessary condition for connectivity but it is not a
sufficient one. Hence, for the connectivity property, the results
in \cite{YiWanLinHuang} imply only the zero-laws under the
scalings (\ref{eq:conjecture_gupta_kumar}) and
(\ref{eq:conjecture_OY}) (and hence under
(\ref{eq:conjecture_gupta_kumar_weak}) and
(\ref{eq:conjecture_OY_weak})), leaving the conjectured one-laws
under the scalings (\ref{eq:conjecture_gupta_kumar}) and
(\ref{eq:conjecture_OY}) open. The less stronger forms of the
conjectured zero-one laws are also open as there exists no results
for the connectivity of the resulting graphs when $1<c \leq 8$
under the scalings (\ref{eq:conjecture_gupta_kumar_weak}) and
(\ref{eq:conjecture_OY_weak}).

\subsection{Contributions}

In this paper, we do not attempt to establish either one of the
conjectures (\ref{eq:conjecture_gupta_kumar_weak}) and
(\ref{eq:conjecture_OY_weak}). Yet, we still would like to
establish a precise characterization of the connectivity
properties of the EG scheme without the full visibility
assumption. With this aim, we study the connectivity properties of
the EG scheme under a simple communication model where channels
are mutually independent, and are either on or off. This amounts
to an overall system model constructed by {\em intersecting} the
random key graph with an Erd\H{o}s-R\'enyi (ER) graph
\cite{Bollobas}. For this random graph structure, we establish
zero-one laws for two basic (and related) graph properties, namely
graph connectivity and the absence of isolated nodes, as the model
parameters are scaled with the number of users -- We identify the
critical thresholds and show that they coincide. Namely, with the
notation introduced so far, we show that if $\alpha$ and $\beta$
are scaled with the number of nodes $n$ and it holds that
\begin{equation}
\alpha_n \beta_n = c \frac{\log n}{n} \label{eq:shown}
\end{equation}
then, the resulting random graph is connected (and has no isolated
nodes) with probability approaching to $1$ (resp. $0$) if $c>1$
(resp. $c<1$); see Section \ref{sec:MainResults} for precise
statements of the results. To the best of our knowledge, these
{\em full} zero-one laws constitute the first {\em complete}
analysis of the EG scheme under {\em non}-full visibility.

Although the communication model considered here may be deemed
simplistic, it does permit a complete analysis of the issues of
interest with the results providing a {\em precise} guideline for
ensuring the secure connectivity of a WSN. Obtaining such precise
guidelines by means of determining the exact threshold of secure
connectivity is particularly crucial in a WSN setting due to a
number of reasons: First, to increase the chances of connectivity,
it is often required to increase the number of keys kept in each
sensor's memory. However, since sensor nodes are expected to have
very limited memory, it is desirable for practical key
distribution schemes to have low memory requirements
\cite{DuDengHanVarshney}. Second, in the EG scheme, there is a
well known \cite{DiPietroManciniMeiPanconesiRadhakrishnan2008}
trade-off between security and connectivity meaning that the more
connected is the network the less secure it is. These point out
the importance of the full zero-one laws established here in
dimensioning the EG scheme as compared to the existing results
\cite{DiPietroMeiManciniPanconesiRadhakrishnan2004,KrzywdziRybarczyk},
where there is a significant gap between the conditions of the
zero-law ($c<1$) and the one-law ($c>8$).

Finally, simulations suggest that the connectivity behavior of the
EG scheme under the on/off channel model is asymptotically
equivalent to that of the EG scheme under the disk model. This
suggests that the zero-one laws obtained here can be taken as an
indication of the validity of the conjectured zero-one law under
the scaling (\ref{eq:conjecture_OY_weak}).

\subsection{Notation and convention}

A word on notation and conventions in use: All limiting
statements, including asymptotic equivalences, are understood with
the number of sensor nodes $n$ going to infinity. The random
variables (rvs) under consideration are all defined on the same
probability triple $(\Omega, {\cal F}, \mathbb{P})$. Probabilistic
statements are made with respect to this probability measure
$\mathbb{P}$, and we denote the corresponding expectation operator
by $\mathbb{E}$. Also, we use the notation $=_{st}$ to indicate
distributional equality. The indicator function of an event $E$ is
denoted by $\1{E}$.  We say that an even holds {\em with high
probability} (whp) if it holds with probability $1$ as $n \to
\infty$. For any discrete set $S$ we write $|S|$ for its
cardinality.

\subsection{Structure of the paper}

The rest of the paper is organized as follows: In Section
\ref{sec:Model}, we give precise definitions and implementation
details of the EG scheme along with a description of the model of
interest. The main results of the paper, namely Theorem
\ref{thm:OneLaw+NodeIsolation} and Theorem
\ref{thm:OneLaw+Connectivity}, are presented in Section
\ref{sec:MainResults} with an extensive simulation results given
in Section \ref{sec:Numerical}. The remaining sections, namely
Sections \ref{sec:Preliminary} through \ref{sec:Last_Parts_3}, are
devoted to establishing the main results of the paper.

\section{The model}
\label{sec:Model}

Under full visibility, the random key distribution scheme of
Eschenauer and Gligor gives rise to a class of random graphs
usually known as random key graphs
\cite{YaganMakowskiConnectivity}; some authors
\cite{BlackburnGerke,Rybarczyk2009} refer to them as uniform
random intersection graphs. Random key graphs are parametrized by
the number $n$ of nodes, the size $P$ of the key pool and the size
$K$ of each key ring with $K \leq P$. To lighten the notation we
often group the integers $P$ and $K$ into the ordered pair $\theta
\equiv (K,P)$.

For each node $i=1, \ldots , n$, let $K_i (\theta)$ denote the
random set of $K$ distinct keys assigned to node $i$. We can think
of $K_i(\theta)$ as an $\mathcal{P}_{K} $-valued rv where
$\mathcal{P}_{K} $ denotes the collection of all subsets of $\{ 1,
\ldots , P \}$ which contain exactly $K$ elements -- Obviously, we
have $|\mathcal{P}_{K} | =  {P \choose K}$. The rvs $K_1(\theta),
\ldots , K_n(\theta)$ are assumed to be {\em i.i.d.} rvs, each of
which is {\em uniformly} distributed over $\mathcal{P}_{K}$ with
\begin{equation}
\bP{ K_i(\theta) = S } = {P \choose K} ^{-1}, \quad S \in
\mathcal{P}_{K} \label{eq:KeyDistrbution1}
\end{equation}
for all $i=1, \ldots , n$. This corresponds to selecting keys
randomly and {\em without} replacement from the key pool.

Distinct nodes $i,j=1, \ldots , n$ are said to be K-adjacent,
written $i \sim_K j$, if they share at least one key in their key
rings, namely
\begin{equation}
i \sim_K j \quad \mbox{iff} \quad K_i(\theta) \cap K_j(\theta)
\neq \emptyset . \label{eq:Adjacency}
\end{equation}
For distinct $i,j =1, \ldots , n$, it is a simple matter to check
that
\begin{equation}
\bP{ K_i (\theta) \cap K_j (\theta) = \emptyset } = q (\theta)
\end{equation}
with
\begin{equation}
q (\theta) = \left \{
\begin{array}{ll}
0 & \mbox{if~ $P <2K$} \\
  &                 \\
\frac{{P-K \choose K}}{{P \choose K}} & \mbox{if~ $2K \leq P$,}
\end{array}
\right . \label{eq:q_theta}
\end{equation}
whence the probability of edge occurrence between any two nodes is
equal to $1-q(\theta)$. The expression (\ref{eq:q_theta}) and
others given later are simple consequences of the often used fact
that
\begin{equation}
\bP{ S \cap K_i(\theta) = \emptyset } = \frac{{P- |S| \choose
K}}{{P \choose K}}, \quad i=1, \ldots ,n
\label{eq:Probab_key_ring_does_not_intersect_S}
\end{equation}
for every subset $S$ of $\{ 1, \ldots , P \}$ with $|S| \leq P-K$.

 With $n=2,3,
\ldots $ and positive integers $K < P$, the adjacency notion
(\ref{eq:Adjacency}) defines the {\em random key graph}
$\mathbb{K}(n; \theta )$ on the vertex set $\{ 1, \ldots , n \}$.

As mentioned earlier, in this paper we seek to account for the
possibility that communication links between nodes may not be
available. To study such situations, we assume a communication
model that consists of independent channels each of which can be
either on or off. Thus, with $\alpha$ in $(0,1)$, let
$\{B_{ij}(\alpha), 1 \leq i < j \leq n\}$ denote i.i.d. $\{0,
1\}$-valued rvs with success probability $\alpha$. The channel
between nodes $i$ and $j$ is available (resp. up) with probability
$\alpha$ and unavailable (resp. down) with the complementary
probability $1-\alpha$.

Distinct nodes $i$ and $j$ are said to be B-adjacent, written $i
\sim_{B} j$, if $B_{ij}(\alpha) = 1$. The notion of B-adjacency
defines the standard Erd\H{o}s-R\'enyi graph
$\mathbb{G}(n;\alpha)$ on the vertex set $\{ 1, \ldots , n \}$.
Obviously,
\[
\bP{ i \sim j}_{B} = \alpha.
\]

The random graph model studied here is obtained by {\em
intersecting} the random key graph $\mathbb{K}(n;\theta)$ with the
ER graph $\mathbb{G}(n;\alpha)$. More precisely, the distinct
nodes $i$ and $j$ are said to be adjacent, written $i \sim j$, if
and only if they are both K-adjacent and B-adjacent, namely
\begin{equation}
i \sim j \quad \mbox{iff} \quad
\begin{array}{c}
K_{i}(\theta) \cap K_{j}(\theta) \neq \emptyset \\
\mbox{and} \\
B_{ij}(\alpha)=1.\\
\end{array}
\label{eq:Adjacency_Intersection}
\end{equation}
The resulting {\em undirected} random graph defined on the vertex
set $\{1, \ldots, n\}$ through this notion of adjacency is denoted
$\mathbb{K\cap G}(n;\theta, \alpha)$.

Throughout the collections of rvs $\{ K_{1}(\theta), \ldots ,
K_{n}(\theta) \}$ and $\{B_{ij}(\alpha), 1 \leq i < j \leq n\}$
are assumed to be independent, in which case the edge occurrence
probability in $\mathbb{K\cap G}(n;\theta,\alpha)$ is given by
\begin{equation}
\bP{i \sim j } = \alpha \cdot \bP{i \sim_{K} j } = \alpha
(1-q(\theta)).
 \label{eq:edge_prob_intersectioN_graph}
\end{equation}

\section{Main results}
\label{sec:MainResults}

To fix the terminology, we refer to any mapping $K,P: \mathbb{N}_0
\rightarrow \mathbb{N}_0$ as a {\em scaling} (for random key
graphs) provided it satisfies the natural conditions
\begin{equation}
K_n \leq P_n, \quad n=1,2, \ldots . \label{eq:ScalingDefn}
\end{equation}
Similarly, any mapping $\alpha: \mathbb{N}_0 \rightarrow (0,1)$
defines a scaling for ER graphs. Finally, a scaling $K,P:
\mathbb{N}_0 \rightarrow \mathbb{N}_0$ is said to be {\em
admissible} if
\begin{equation}
2 \leq  K_n \label{eq:AdmissibilityA}
\end{equation}
for {\em all} $n=1,2, \ldots $ {\em sufficiently} large.

To lighten the notation we often group the parameters $K$, $P$ and
$\alpha$ into the ordered triple $\Theta \equiv (K,P,\alpha)=
(\theta, \alpha)$. Hence, a mapping $\Theta: \mathbb{N}_0
\rightarrow \mathbb{N}_0 \times \mathbb{N}_0 \times (0,1)$ defines
a scaling for the intersection graph $\mathbb{K\cap G}(n;\Theta)$
provided the condition (\ref{eq:ScalingDefn}) holds.

\subsection{Absence of isolated nodes}
\label{subsec:ResultsAbsenceIsolatedNodes}

The first result gives a zero-one law for the absence of isolated
nodes.

\begin{theorem}
{\sl Consider an admissible scaling $K,P: \mathbb{N}_0 \rightarrow
\mathbb{N}_0$ and a scaling $\alpha: \mathbb{N}_0 \rightarrow
(0,1)$ such that
\begin{equation}
\alpha_n (1-q(\theta_n)) \sim c \frac{\log n}{n}, \quad n=1,2,
\ldots \label{eq:scalinglaw}
\end{equation}
for some $c>0$. If $\lim_{n \to \infty}\alpha_n \log n =
\alpha^\star$ exists, then we have
\begin{eqnarray}
\lim_{n \rightarrow \infty } \bP{
\begin{array}{c}
\mathbb{K \cap G}(n;\Theta_n)~\mbox{contains} \\
\mbox{~no~isolated~nodes} \\
\end{array}
} = \left \{
\begin{array}{ll}
0 & \mbox{if~ $c <1$} \\
  &                      \\
1 & \mbox{if~$c > 1$.}
\end{array}
\right . \label{eq:OneLaw+NodeIsolation}
\end{eqnarray}
} \label{thm:OneLaw+NodeIsolation}
\end{theorem}

The condition (\ref{eq:scalinglaw}) on the scalings will often be
used in the equivalent form
\begin{equation}
\alpha_n (1-q(\theta_n)) = c_n \frac{\log n}{n}, \quad n=1,2,
\ldots \label{eq:scalinglawEquivalent}
\end{equation}
with the sequence $c: \mathbb{N}_0 \rightarrow \mathbb{R}_+$
satisfying $\lim_{n \rightarrow \infty} c_n = c$.

The assumption that $\lim_{n \to \infty}\alpha_n \log n =
\alpha^\star$ exists is made due to technical reasons and it is
much weaker than the condition $\lim_{n \to \infty}\alpha_n \log
n=\infty$ assumed in \cite{YiWanLinHuang}.

\subsection{Connectivity}
\label{subsec:ResultsConnectivity}

An analog of Theorem \ref{thm:OneLaw+NodeIsolation} also holds for
the property of graph connectivity.

\begin{theorem}
{\sl Consider an admissible scaling $K,P: \mathbb{N}_0 \rightarrow
\mathbb{N}_0$ and a scaling $\alpha: \mathbb{N}_0 \rightarrow
(0,1)$ such that (\ref{eq:scalinglaw}) holds for some $c>0$. If
$\lim_{n \to \infty}\alpha_n \log n = \alpha^\star$ exists then we
have
\begin{eqnarray}
 \lim_{n \rightarrow \infty } \bP{\mathbb{K \cap
G}(n;\Theta_n)~\mbox{ is connected} } = 0 \quad \mbox{if~ $c <1$}
\label{eq:ZeroLaw+Connectivity}
\end{eqnarray}
On the other hand, if there exists some $\sigma>0$ such that
\begin{equation}
\sigma n \leq P_n \label{eq:OneLaw+ConnectivityExtraCondition}
\end{equation}
for all $n=1,2, \ldots $ sufficiently large, then we have
\begin{eqnarray}
 \lim_{n \rightarrow \infty } \bP{\mathbb{K \cap
G}(n;\Theta_n)~\mbox{ is connected} } = 1 \quad \mbox{if~ $c>1.$}
\label{eq:OneLaw+Connectivity}
\end{eqnarray}
 } \label{thm:OneLaw+Connectivity}
\end{theorem}

Comparing Theorem \ref{thm:OneLaw+Connectivity} with Theorem
\ref{thm:OneLaw+NodeIsolation}, we see that the class of random
graphs studied here provides one more instance where the zero-one
laws for absence of isolated nodes and connectivity coincide, viz.
ER graphs \cite{Bollobas}, random geometric graphs
\cite{PenroseBook} or random key graphs
\cite{BlackburnGerke,Rybarczyk2009,YaganMakowskiConnectivity}.

The condition (\ref{eq:OneLaw+ConnectivityExtraCondition}) states
that the size of the key pool $P_n$ should grow at least linearly
with the number of sensor nodes in the network. Although this
condition is enforced merely for technical reasons, it is not at
all a stringent constraint in a realistic WSN scenario. In fact,
it holds trivially for any realization as it is expected
\cite{DiPietroMeiManciniPanconesiRadhakrishnan2004,EschenauerGligor}
that the size of the key pool will be much larger than the number
of participating nodes for security purposes.

Theorem \ref{thm:OneLaw+Connectivity} cannot hold if the condition
(\ref{eq:AdmissibilityA}) fails. This is a simple consequence of
the fact that if $K_n=1$ for all $n$ sufficiently large, than the
random key graph $\mathbb{K}(n;\theta)$ is disconnected with high
probability unless it also holds that $P_n=1$ for all $n$
sufficiently large; see \cite[Lemma 7.1.2, pp. 99]{YaganThesis}.


\section{Numerical Results}
\label{sec:Numerical}

We now present numerical results and simulations that show the
validity of Theorem \ref{thm:OneLaw+NodeIsolation} and Theorem
\ref{thm:OneLaw+Connectivity}.

In all experiments, we fix the number of nodes at $n=500$ and the
size of the key pool at $P=10,000$. We consider the channel
parameters $\alpha=0.2$, $\alpha=0.4$, $\alpha=0.6$ and
$\alpha=0.8$, while varying the parameter $K$ from $1$ to $35$.
For each parameter pair $(K,\alpha)$, we generate $200$
independent samples of the graph $\mathbb{K} \cap
\mathbb{G}(n;K,P,\alpha)$ and count the number of times (out of a
possible 200) that the obtained graphs i) have no isolated nodes
and ii) are connected. Dividing the counts by $200$, we obtain the
(empirical) probabilities for the events of interest. In all
cases, we observe that $\mathbb{K} \cap \mathbb{G}(n;K,P,\alpha)$
is connected whenever it has no isolated nodes yielding the same
empirical probability for both events. This confirms the
asymptotic equivalence of the connectivity and absence of isolated
nodes properties in $\mathbb{K} \cap \mathbb{G}(n;\Theta)$ as
stated in Proposition \ref{prop:OneLawAfterReduction}.

\begin{figure}[!t]
\vspace{-1cm}
 \hspace{-0.75 cm}
\includegraphics[totalheight=0.37\textheight,
width=0.55\textwidth]{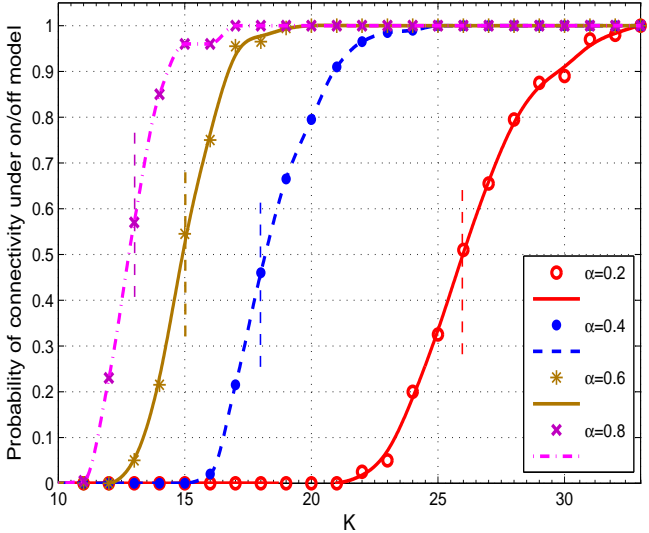} \caption{Empirical probability
that $\mathbb{K} \cap \mathbb{G}(n;K,P,\alpha)$
         is connected as a function of $K$ for
         $\alpha=0.2$, $\alpha=0.4$, $\alpha=0.6$, $\alpha=0.8$
         with $n=500$ and $P=10,000$; in each case, the empirical probability value is
obtained by averaging over $200$ experiments. Vertical dashed
lines stand for the critical threshold of connectivity asserted by
Theorem \ref{thm:OneLaw+Connectivity}. It is clear that the
theoretical findings are in perfect agreement with the
experimental observations. } \label{figure:connect}
\end{figure}

In Figure \ref{figure:connect}, we depict the resulting empirical
probability of connectivity in $\mathbb{K} \cap
\mathbb{G}(n;K,P,\alpha)$ versus $K$ for several $\alpha$ values.
For a better visualization of the data, we use the curve fitting
tool of MATLAB. For each $\alpha$ value, we show the critical
threshold of connectivity asserted by Theorem
\ref{thm:OneLaw+Connectivity} by a vertical dashed line. Namely,
the vertical dashed lines stand for the minimum integer value of
$K$ that satisfies
\begin{equation}
1-q(\theta) = 1-{{{P-K} \choose K} \over {P \choose K}} >
\frac{1}{\alpha}\frac{\log n}{n}. \label{eq:threshold}
\end{equation}
Even with $n=500$, the threshold behavior of the probability of
connectivity is evident from the plots. Of course, as $n$ gets
large, we expect the curves to look more like a {\em shifted unit
step} function with a jump discontinuity (i.e., a threshold) at
around the $K$ value that gives
$\bP{\textrm{Connectivity}}=\frac{1}{2}$ in the current plots.
Thus, for each value of $\alpha$, we see that the connectivity
threshold prescribed by (\ref{eq:threshold}) is in perfect
agreement with the experimentally observed threshold of
connectivity.

\begin{figure}[!t]
\vspace{-1cm}
 \hspace{-0.65 cm}
\includegraphics[totalheight=0.37\textheight,
width=0.55\textwidth]{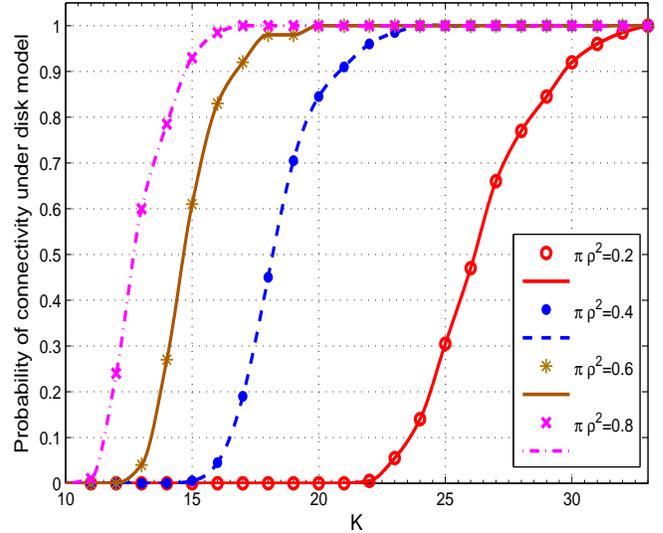} \caption{Empirical probability
that $\mathbb{K} \cap \mathbb{H}(n;K,P,\rho)$
         is connected as a function of $K$. The number of nodes is set
         to $n=500$ and we take $P=10,000$. The resemblance of the plots
         to
         those of Figure
\ref{figure:connect} suggests that the connectivity behaviors of
the models $\mathbb{K\cap G}(n;K,P,\alpha)$ and $\mathbb{K\cap
H}(n;K,P,\rho)$ are quite similar under the matching condition
$\pi \rho^2 =\alpha$. } \label{figure:disk}
\end{figure}

One possible extension of the work presented here would be to
consider a more realistic communication model; e.g., the popular
disk model \cite{GuptaKumar} instead of the on/off channel model.
As discussed in the Introduction, the disk model induces random
geometric graphs \cite{PenroseBook} denoted by
$\mathbb{H}(n;\rho)$, where $n$ is the number of nodes and $\rho$
is the transmission range. Under the disk model, studying the EG
scheme amounts to analyzing the intersection of
$\mathbb{K}(n;\theta)$ and $\mathbb{H}(n;\rho)$, say
$\mathbb{K\cap H}(n;K,P,\rho)$. To compare the connectivity
behavior of the EG scheme under the disk model with that of the
on-off channel model, consider $200$ nodes distributed uniformly
and independently over a folded unit square $[0,1]^2$ with
toroidal (continuous) boundary conditions. Since there are no
border effects, it is easy to check that
\[
\bP{\: \parallel \boldsymbol{x_i} -\boldsymbol{x_j} \parallel<\rho
\:} = \pi \rho ^2, \quad i \neq j, \:\: i,j=1,2, \ldots, n.
\]
whenever $\rho<0.5$. Thus, we can match the two communication
models $\mathbb{G}(n;\alpha)$ and $\mathbb{H}(n;\rho)$ by
requiring $\pi \rho^2 = \alpha$. Using the same procedure that
produced Figure \ref{figure:connect}, we obtain the empirical
probability that $\mathbb{K\cap H}(n;K,P,\rho)$ is connected
versus $K$ for various $\rho$ values. The results are depicted in
Figure \ref{figure:disk} whose resemblance with Figure
\ref{figure:connect} suggests that the connectivity behaviors of
the models $\mathbb{K\cap G}(n;K,P,\alpha)$ and $\mathbb{K \cap
H}(n;K,P,\rho)$ are quite similar under the matching condition
$\pi \rho^2 =\alpha$. This raises the possibility that the results
obtained here for the on/off communication model can be taken as
an indication of the validity of the conjectured zero-one law by
Gupta and Kumar \cite{GuptaKumar} given under the scaling
(\ref{eq:conjecture_gupta_kumar_weak}).

\section{Preliminaries} \label{sec:Preliminary}

Before we give a proof of Theorem \ref{thm:OneLaw+NodeIsolation}
and Theorem \ref{thm:OneLaw+Connectivity}, we give some
preliminary results that will help establish them.

The following inequality will prove useful in a number of places.
\begin{lemma}
{\sl For any $\theta=(K,P)$ with positive integers $K,P$ and any
scalar $a \geq 1$, we have
\begin{equation}
\frac{{P-\lceil a K \rceil \choose K}}{{P \choose K}} \leq
q(\theta) ^ a \label{eq:prelimiary}
\end{equation}
}
\end{lemma}

\myproof Observe that under the enforced assumptions it always
holds that $q(\theta)\geq0$, so that (\ref{eq:prelimiary}) holds
trivially if $K + \lceil a K \rceil > P $. On the other hand, if
$K + \lceil a K \rceil \leq P $, we can use the relation
\cite[Lemma 5.4.1, pp. 79]{YaganThesis}
\begin{eqnarray}\nonumber
\frac{{P- L \choose K}}{{P \choose K}} &=& \prod_{\ell=0}^{K-1}
\left ( 1 - \frac{L}{P-\ell} \right )
\end{eqnarray}
valid for any $L$ such that $L+K \leq P$. Substituting we find
\begin{eqnarray}\nonumber
\frac{{P-  \lceil a K \rceil \choose K}}{{P \choose K}} &=&
\prod_{\ell=0}^{K-1} \left ( 1 - \frac{ \lceil a K \rceil}{P-\ell}
\right )
\\ \label{eq:prelim_1}
&\leq& \prod_{\ell=0}^{K-1} \left ( 1 - \frac{ a K }{P-\ell}
\right )
\end{eqnarray}
and
\begin{equation}
q(\theta) = \prod_{\ell=0}^{K-1} \left ( 1 - \frac{ K }{P-\ell}
\right ) \label{eq:prelim_2}
\end{equation}

In view of (\ref{eq:prelim_1}) and (\ref{eq:prelim_2}), the
desired inequality (\ref{eq:prelimiary}) will follow if we show
that
\begin{equation}
 1 - \frac{ a K }{P-\ell} \leq \left ( 1 - \frac{ K }{P-\ell}
\right ) ^ a, \quad \ell=0,1,\ldots, K-1 \label{eq:prelim_to_show}
\end{equation}
For each $\ell =0,1,\ldots, K-1$, this is immediate once we note
that
\[
1 - \left ( 1 - \frac{K}{P-\ell} \right )^{a} = \int_{ 1 -
\frac{K}{P-\ell} }^1 a t^{a-1}dt \leq \frac{a K}{P-\ell}
\]
by a crude bounding argument and  (\ref{eq:prelimiary}) follows.
\myendpf

We also find it useful to make use of the next result:
\begin{lemma}
{\sl Consider $\theta=(K,P)$ with positive integers $K,P$ such
that $K \leq 2 P$. For any $0< \lambda < 1$, we have
\begin{equation}
1-q(\theta) ^ \lambda \geq \lambda
(1-q(\theta))\label{eq:prelimiaryB}
\end{equation}
}
\end{lemma}

\myproof

Since $\lambda < 1$, we have
\[
1-q(\theta) ^ \lambda = \int_{q(\theta) }^1 \lambda
t^{\lambda-1}dt \geq \int_{q(\theta) }^1 \lambda dt = \lambda
(1-q(\theta)).
\]
\myendpf

In the proof of Theorem \ref{thm:OneLaw+Connectivity}, we will
make use of a result that is a direct consequence of the condition
(\ref{eq:OneLaw+ConnectivityExtraCondition}). 

\begin{lemma}
{\sl Consider a scaling $\Theta: \mathbb{N}_0 \rightarrow
\mathbb{N}_0 \times \mathbb{N}_0 \times (0,1) $ such that
(\ref{eq:scalinglaw}) holds for some $c>0$ and
(\ref{eq:OneLaw+ConnectivityExtraCondition}) holds for some
$\sigma>0$. Then, for any $\delta>0$, there exists a finite
integer $n^\star=n^\star(\delta)$ such that for all $n \geq
n^\star$ sufficiently large, we have
\begin{equation}
K_n \geq \sqrt{\left(1-\delta\right)\sigma c \log n}
\label{eq:usefulcons3}
\end{equation}
for all $n \geq n^\star$ sufficiently large.}
\label{lem:usefulcons2}
\end{lemma}

\myproof Under the enforced assumptions it can be seen \cite[Lemma
7.4.3, pp. 118]{YaganThesis} from (\ref{eq:prelim_2}) that
\begin{equation}
1 - q (\theta_n) \leq \frac{K^2_n}{P_n-K_n}.
\label{eq:Equivalence1}
\end{equation}
Multiplying the above inequality by $\alpha_n$ and using the
scaling condition $(\ref{eq:scalinglaw})$, we find
\[
c_n \frac{\log n}{n} \leq \alpha_n\frac{K^2_n}{P_n}
\frac{1}{1-\frac{K_n}{P_n}},
\]
or equivalently
\[
\alpha_n\frac{K^2_n}{P_n} \geq c_n \frac{\log n}{n}
\left(1-\frac{K_n}{P_n}\right)
\]
 where the sequence $c: \mathbb{N}_0
\rightarrow \mathbb{R}_+$ satisfies $\lim_{n \rightarrow \infty}
c_n = c$. Invoking (\ref{eq:OneLaw+ConnectivityExtraCondition}),
we get
\[
K_n^2 \geq \frac{1}{\alpha_n} c_n \sigma  \log n
\left(1-\frac{K_n}{P_n}\right) \geq c_n \sigma \log n
\left(1-\frac{K_n}{\sigma n}\right)
\]
upon using the fact that $\alpha_n \leq 1$ for each $n=1,2,
\ldots.$ This is equivalent to having
\begin{eqnarray}\nonumber
K_n^2 + K_n \frac{c_n \log n}{n} \geq c_n \sigma \log n
\end{eqnarray}
which yields
\[
K_n^2 (1+o(1)) \geq  c_n \sigma \log n
\]
The desired conclusion (\ref{eq:usefulcons3}) is now immediate.
\myendpf

\section{A proof of Theorem \ref{thm:OneLaw+NodeIsolation}}
\label{sec:ProofTheoremNodeIsolation}

We prove Theorem  \ref{thm:OneLaw+NodeIsolation} by the method of
first and second moments \cite[p.  55]{JansonLuczakRucinski}
applied to the total number of isolated nodes in $\mathbb{K \cap
G}(n;\Theta)$. First some notation: Fix $n=2,3, \ldots $ and
consider $\Theta = (K,P,\alpha)$ with $\alpha$ in $(0,1)$ and
positive integers $K,P$ such that $K \leq P$. With
\[
\chi_{n,i}(\Theta) := \1{ {\rm Node~}i~{\rm is~isolated~in~}
                         \mathbb{K \cap G}(n;\Theta) }
\]
for each $i=1, \ldots , n$, the number of isolated nodes in
$\mathbb{K \cap G}(n;\Theta)$ is simply given by
\[
I_n (\Theta) := \sum_{i=1}^n \chi_{n,i}(\Theta).
\]
The random graph $\mathbb{K \cap G}(n;\Theta)$ has no isolated
nodes if and only if $I_n (\Theta) = 0$.

The method of first moment \cite[Eqn (3.10), p.
55]{JansonLuczakRucinski} relies on the well-known bound
\begin{equation}
1 - \bE{ I_n (\Theta) } \leq \bP{  I_n (\Theta) = 0 }
\label{eq:FirstMoment}
\end{equation}
while the method of second moment \cite[Remark 3.1, p.
55]{JansonLuczakRucinski} has its starting point in the inequality
\begin{equation}
\bP{  I_n (\Theta) = 0 } \leq 1 - \frac{ \bE{ I_n (\Theta)}^2}{
\bE{ I_n (\Theta) ^2} }. \label{eq:SecondMoment}
\end{equation}

The rvs $\chi_{n,1}(\Theta), \ldots , \chi_{n,n} (\Theta)$ being
exchangeable, we find
\begin{equation}
\bE{ I_n (\Theta)} = n \bE{ \chi_{n,1} (\Theta) }
\label{eq:FirstMomentExpression}
\end{equation}
and
\begin{eqnarray}\label{eq:SecondMomentExpression}
\bE{ I_n (\Theta)^2 } = n \bE{ \chi_{n,1} (\Theta) }
 +  n(n-1) \bE{ \chi_{n,1} (\Theta)  \chi_{n,2} (\Theta) } \nonumber
\end{eqnarray}
by the binary nature of the rvs involved. It then follows that
\begin{eqnarray}
\frac{ \bE{ I_n (\Theta)^2 }}{ \bE{ I_n (\Theta) }^2 } &=& \frac{
1}{ n\bE{ \chi_{n,1} (\Theta) } }
\nonumber \\
& & + \frac{n-1}{n} \cdot \frac{\bE{ \chi_{n,1} (\Theta)
\chi_{n,2} (\Theta) }}
     {\left (  \bE{ \chi_{n,1} (\Theta) } \right )^2 }.
\label{eq:SecondMomentRatio}
\end{eqnarray}

From (\ref{eq:FirstMoment}) and (\ref{eq:FirstMomentExpression})
we see that the one-law $\lim_{n\to \infty} \bP{I_n(\Theta_n) = 0}
= 1$ will be established if we show that
\begin{equation}
\lim_{n \to \infty} n \bE{ \chi_{n,1} (\Theta_n) }= 0.
\label{eq:OneLaw+NodeIsolation+convergence}
\end{equation}
It is also plain from (\ref{eq:SecondMoment}) and
(\ref{eq:SecondMomentExpression}) that the zero-law $\lim_{n \to
\infty} \bP{I_n(\Theta_n) = 0} = 0$ holds if
\begin{equation}
\lim_{n \to \infty} n \bE{ \chi_{n,1} (\Theta_n) }= \infty
\label{eq:OneLaw+NodeIsolation+convergence2}
\end{equation}
and
\begin{equation}
\limsup_{n \to \infty} \left( \frac{\bE{ \chi_{n,1} (\Theta_n)
\chi_{n,2} (\Theta_n) }}
     {\left (  \bE{ \chi_{n,1} (\Theta_n) } \right )^2 }
\right) \leq 1. \label{eq:ZeroLaw+NodeIsolation+convergence}
\end{equation}

The proof of Theorem \ref{thm:OneLaw+NodeIsolation} passes through
the next two technical propositions which establish
(\ref{eq:OneLaw+NodeIsolation+convergence}),
(\ref{eq:OneLaw+NodeIsolation+convergence2}) and
(\ref{eq:ZeroLaw+NodeIsolation+convergence}) under the appropriate
conditions on the scaling $\Theta: \mathbb{N}_0 \rightarrow
\mathbb{N}_0 \times \mathbb{N}_0 \times (0,1)$.

\begin{proposition}
{\sl Consider a scaling $K,P: \mathbb{N}_0 \rightarrow
\mathbb{N}_0$ and a scaling $\alpha: \mathbb{N}_0 \rightarrow
(0,1)$ such that (\ref{eq:scalinglaw}) holds for some $c>0$. Then,
we have
\begin{equation}
\lim_{n \rightarrow \infty } n\bE{ \chi_{n,1} (\Theta_n) } = \left
\{
\begin{array}{ll}
0 & \mbox{if~ $c > 1$} \\
  &                      \\
\infty & \mbox{if~$c < 1$}
\end{array}
\right . \label{eq:NodeIsolation+FirstMoment}
\end{equation}
} \label{prop:Technical1}
\end{proposition}
A proof of Proposition \ref{prop:Technical1} is given in Section
\ref{sec:ProofPropositionTechnical1}.

\begin{proposition}
{\sl Consider an admissible scaling $K,P: \mathbb{N}_0 \rightarrow
\mathbb{N}_0$ and a scaling $\alpha: \mathbb{N}_0 \rightarrow
(0,1)$ such that (\ref{eq:scalinglaw}) holds for some $c>0$. If
$\lim_{n \to \infty}\alpha_n \log n =\alpha^\star$ exists, then we
have (\ref{eq:ZeroLaw+NodeIsolation+convergence}) whenever $c <
1$. } \label{prop:Technical2}
\end{proposition}
A proof of Proposition \ref{prop:Technical2} can be found in
Section \ref{sec:ProofPropositionTechnical2}.

To complete the proof of Theorem \ref{thm:OneLaw+NodeIsolation},
pick a scaling $\Theta: \mathbb{N}_0 \rightarrow \mathbb{N}_0
\times \mathbb{N}_0 \times (0,1) $ such that (\ref{eq:scalinglaw})
holds for some $c>0$ and $\lim_{n \to
\infty}\alpha_n=\alpha^\star$ exists. Under the condition $c> 1$
we get (\ref{eq:OneLaw+NodeIsolation+convergence}) from
Proposition \ref{prop:Technical1}, and the one-law $\lim_{n\to
\infty} \bP{I_n(\Theta_n) = 0} = 1$ follows. Next, assume that $c<
1$. We obtain (\ref{eq:OneLaw+NodeIsolation+convergence2}) and
(\ref{eq:ZeroLaw+NodeIsolation+convergence}) with the help of
Propositions \ref{prop:Technical1} and \ref{prop:Technical2},
respectively. The conclusion $\lim_{n \to \infty}
\bP{I_n(\Theta_n) = 0} = 0$ follows and Theorem
\ref{thm:OneLaw+NodeIsolation} is now established. \myendpf

\subsection{A proof of Proposition \ref{prop:Technical1}}
\label{sec:ProofPropositionTechnical1}

In the course of proving Proposition \ref{prop:Technical1} we make
use of the decomposition
\begin{equation}
\log ( 1 - x ) = -x - \Psi (x), \quad 0 \leq x < 1
\label{eq:LogDecomposition}
\end{equation}
with
\[
\Psi(x) := \int_0^x \frac{t}{1-t} dt
\]
on that range. Note that
\[
\lim_{x \downarrow 0} \frac{ \Psi(x) }{x^2} = \frac{1}{2}.
\]

Fix $n=2,3, \ldots $ and consider $\Theta = (K,P,\alpha)$ with
$\alpha$ in $(0,1)$ and positive integers $K,P$ such that $K \leq
P$. It is easy to see that $\chi_{n,1} (\Theta)=1$, meaning that
node $1$ is isolated, if and only if
\[
B_{1j}(\alpha)=0~\vee~K_1(\theta) \cap K_j(\theta) =
\emptyset,\quad j=2,\ldots,n.
\]
Conditioning on $K_1(\theta)$, we get by independence that
\begin{equation}\label{eq:Prob_of_isol}
\bE{ \chi_{n,1} (\Theta) } =
\left(1-\alpha(1-q(\theta))\right)^{n-1}.
\end{equation}

Now consider a scaling $\Theta: \mathbb{N}_0 \rightarrow
\mathbb{N}_0 \times \mathbb{N}_0 \times (0,1)$ such that
(\ref{eq:scalinglaw}) holds for some $c>0$ and replace $\Theta$ by
$\Theta_n$ in (\ref{eq:Prob_of_isol}) according to this scaling.
Using decomposition (\ref{eq:LogDecomposition}) we get
\begin{equation}
n \bE{ \chi_{n,1} (\Theta_n) } = e^{\beta_n}
\label{eq:FirstMoment_Exp2}
\end{equation}
with
\begin{eqnarray}
\lefteqn{\beta_n} &&
\\ \nonumber
&=& \log n - (n-1)\left(\alpha_n(1-q(\theta_n)
+\psi\left(\alpha_n(1-q(\theta_n))\right)\right)
\\ \nonumber
&=& \log n - (n-1)\left(c_n \frac{\log n}{n} +\psi\left(c_n
\frac{\log n}{n}\right)\right)
\\ \nonumber
&=& \log n \left(1-c_n\frac{n-1}{n}\right)
\\ \nonumber
& & ~ -(n-1)\left(c_n \frac{\log n}{n}\right)^2
\frac{\psi\left(c_n \frac{\log n}{n}\right)}{\left(c_n \frac{\log
n}{n}\right)^2}
\end{eqnarray}
where the sequence $c: \mathbb{N}_0 \rightarrow \mathbb{R}_+$
satisfies $\lim_{n \rightarrow \infty} c_n = c$. Let $n$ go to
infinity in this last expression and recall the behavior of
$\Psi(x)$ at $x=0$ mentioned earlier. We have
\[
\lim_{n \to \infty} (n-1)\left(c_n \frac{\log n}{n}\right)^2
\frac{\psi\left(c_n \frac{\log n}{n}\right)}{\left(c_n \frac{\log
n}{n}\right)^2}=0.
\]
Noting also that
\[
\lim_{n \to \infty} \left(1-c_n\frac{n-1}{n}\right) = 1-c,
\]
we get $\lim_{n \to \infty}\beta_n=\infty$ (resp. $-\infty$)
whenever $c<1$ (resp. $c>1$). The desired condition
(\ref{eq:NodeIsolation+FirstMoment}) is now immediate via
(\ref{eq:FirstMoment_Exp2}). \myendpf

\subsection{A proof of
         Proposition \ref{prop:Technical2}}
\label{sec:ProofPropositionTechnical2}

As expected, the first step in proving Proposition
\ref{prop:Technical2} consists in evaluating the cross moment
appearing in the numerator of
(\ref{eq:ZeroLaw+NodeIsolation+convergence}). Fix $n=2,3, \ldots $
and consider $\Theta = (K,P,\alpha)$ with $\alpha$ in $(0,1)$ and
positive integers $K,P$ such that $K \leq P$. Define the
$\{0,1\}$-valued rv $u(\theta)$ by
\begin{equation}
u(\theta) := \1{K_1 (\theta) \cap K_2(\theta) \neq \emptyset}.
\label{eq:u}
\end{equation}
Next, with $r=1,2, \ldots, n-1$ define $v_{r,j}(\alpha)$ by
\begin{eqnarray}
v_{r,j}(\alpha) := \{ i=1,2,\ldots,r : B_{ij}(\alpha) =1 \}
\label{eq:v}
\end{eqnarray}
for each $j=r+1,\ldots,n$. In other words, $v_{r,j}(\alpha)$ is
the set of nodes in $1,\ldots, r$ that have an edge with the node
$j$ in the communication graph $\mathbb{G}(n;\alpha)$.
Conditioning in $K_1(\theta)$ and $K_2(\theta)$, it is now a
simple matter to check that
\begin{eqnarray}\nonumber
\bE{ \chi_{n,1} (\Theta) \chi_{n,2} (\Theta) } = \bE{ (1-\alpha)^{
u(\theta)} \prod_{j=3}^{n}{{{P-|\cup_{i \in v_{2,j}}
K_i(\theta)|}\choose{K}}\over{{P}\choose{K}} }}
\end{eqnarray}

In order to efficiently bound this term, we first observe that
under the event $u(\theta)=0$ (which happens with probability
$q(\theta)$), we have
\[
|\cup_{i \in v_{2,j}(\alpha)} K_i(\theta)| = |v_{2,j}(\alpha)| K,
\quad j=3,\ldots,n
\]
and it is plain by direct inspection and (\ref{eq:prelimiary})
that
\[
{ {{P-|v_{2,j}(\alpha)|K}\choose{K}}\over{{P}\choose{K}}}  \leq
q(\theta)^{|v_{2,j}(\alpha)|}
\]
As a result, we find
\begin{eqnarray}\nonumber
\lefteqn{\bE{ (1-\alpha)^{ u(\theta)} \prod_{j=3}^{n}{{{P-|\cup_{i
\in v_{2,j}(\alpha)} K_i(\theta)|}\choose{K}}\over{{P}\choose{K}}
} \: \Bigg | \: u(\theta)=0}} &&
\\ \nonumber
&\leq&
 \bE{ \prod_{j=3}^{n} q(\theta)^{|v_{2,j}(\alpha)|} }
\\ \nonumber
&=& \bE{q(\theta)^{|v_{2,3}(\alpha)|} }^{n-2}
\\ \nonumber
&=& \left(\sum_{i=0}^{2} {2 \choose i} \alpha^{i} (1-\alpha)^{2-i}
q(\theta)^{i} \right)^{n-2}
\\ \label{eq:secondMoment_u_0}
&=& \left(1-\alpha(1-q(\theta))\right)^{2(n-2)}
\end{eqnarray}
as we note that $\{|v_{r,j}(\alpha)|\}_{j=r+1}^{n}$ are i.i.d.
random variables with
\[
|v_{r,j}(\alpha)| =_{\mbox{st}} \mbox{Bin}(r,\alpha),\quad
j=r+1,\ldots, n.
\]

On the other hand if $u(\theta)=1$ (which happens with probability
$1-q(\theta)$) we get
\begin{eqnarray}
\lefteqn{|\cup_{i \in v_{2,j}(\alpha)} K_i(\theta)|} && \\
\nonumber
 &=& \left \{
\begin{array}{lll}
0 & \mbox{if~$|v_{2,j}(\alpha)|=0$} \\
K & \mbox{if~$|v_{2,j}(\alpha)|=1$}
  &                      \\
2K-|K_1(\theta)\cap K_2(\theta)|
 & \mbox{if~$|v_{2,j}(\alpha)|=2$}
\end{array}
\right .
\end{eqnarray}
for each $j=3,\ldots,n$. Therefore, crude bounding argument gives
\[
|\cup_{i \in v_{2,j}(\alpha)} K_i(\theta)| \geq K
\1{|v_{2,j}(\alpha)|>0}
\]
yielding
\[
{ {{P-|\cup_{i \in v_{2,j}(\alpha)}
K_i(\theta)|}\choose{K}}\over{{P}\choose{K}}}  \leq q(\theta) ^{
\1{|v_{2,j}(\alpha)|>0}}
\]

With these in mind we obtain
\begin{eqnarray}\nonumber
\lefteqn{\bE{ (1-\alpha)^{ u(\theta)} \prod_{j=3}^{n}{{{P-|\cup_{i
\in v_{2,j}(\alpha)} K_i(\theta)|}\choose{K}}\over{{P}\choose{K}}
} \: \Bigg | \: u(\theta)=1}} &&
\\ \nonumber
&\leq&
 (1-\alpha)\bE{ \prod_{j=3}^{n} q(\theta)^{\1{|v_{2,j}(\alpha)|>0}} }
 \\ \nonumber
 &=&  (1-\alpha)\bE{ q(\theta)^{\1{|v_{2,3}(\alpha)|>0}} }^{n-2}
\\ \nonumber
&=& (1-\alpha)\left((1-\alpha)^2 +
\left(1-(1-\alpha)^2\right)q(\theta)\right)^{n-2}
\\ \nonumber
&=&
(1-\alpha)\left(\left(1-\alpha(1-q(\theta))\right)^2+\alpha^2q(\theta)(1-q(\theta))\right)^{n-2}
\\ \label{eq:secondMoment_u_1}
&\leq&
\left(\left(1-\alpha(1-q(\theta))\right)^2+\alpha^2q(\theta)(1-q(\theta))\right)^{n-2}
\end{eqnarray}

Recalling (\ref{eq:Prob_of_isol}), (\ref{eq:secondMoment_u_0}) and
(\ref{eq:secondMoment_u_1}) we find
\begin{eqnarray}\label{eq:SecondMoment_Bound}
\lefteqn{\frac{\bE{ \chi_{n,1} (\Theta) \chi_{n,2} (\Theta) }}
     {\left (  \bE{ \chi_{n,1} (\Theta) } \right )^2 }} &&
     \\ \nonumber
&\leq& q(\theta)\frac{\left(1-\alpha(1-q(\theta))\right)^{2(n-2)}}
    {\left(1-\alpha(1-q(\theta))\right)^{2(n-1)}} +
    (1-q(\theta))\times
    \\\nonumber
    &&~ \times \frac{
\left(\left(1-\alpha(1-q(\theta))\right)^2+\alpha^2q(\theta)(1-q(\theta))\right)^{n-2}}
    {\left(1-\alpha(1-q(\theta))\right)^{2(n-1)}}
    \\\nonumber
    &=&\frac{q(\theta)}{\left(1-\alpha(1-q(\theta))\right)^{2}}
    \\ \nonumber
    && ~ +
    \frac{1-q(\theta)}{\left(1-\alpha(1-q(\theta))\right)^{2}}
    \left(1+\frac{\alpha^2q(\theta)(1-q(\theta))}
    {\left(1-\alpha(1-q(\theta))\right)^{2}}\right)^{n-2}
    \\ \nonumber
    &\leq&
    \frac{q(\theta)+(1-q(\theta))\exp\{\frac{\alpha^2(1-q(\theta))n}
    {\left(1-\alpha(1-q(\theta))\right)^{2}}\}}
    {\left(1-\alpha(1-q(\theta))\right)^{2}}
\end{eqnarray}

Now consider a scaling $\Theta: \mathbb{N}_0 \rightarrow
\mathbb{N}_0 \times \mathbb{N}_0 \times (0,1)$ such that
(\ref{eq:scalinglaw}) holds for some $c<1$ and replace $\Theta$ by
$\Theta_n$ in (\ref{eq:SecondMoment_Bound}) according to this
scaling. Invoking (\ref{eq:scalinglaw}) it is immediate that
\[
\lim_{n \to \infty} \left(1-\alpha_n(1-q(\theta_n))\right)^{2}=1
\]
and the desired condition
(\ref{eq:ZeroLaw+NodeIsolation+convergence}) will follow if we
show that
\begin{equation}\label{eq:condition_to_zero_law}
\limsup_{n \to
\infty}\left(q(\theta_n)+(1-q(\theta_n))\exp\left\{\frac{\alpha_n
c_n \log n}{\left(1-c_n\frac{\log
n}{n}\right)^{2}}\right\}\right)\leq 1
\end{equation}
with $\lim_{n \to \infty}c_n=c<1$. As in the statement of
Proposition \ref{prop:Technical2} assume that $\lim_{n \to \infty}
\alpha_n \log n = \alpha^\star$ exists. We consider the cases
$\alpha^ \star = 0$ and $\alpha^\star \in (0,\infty]$ separately.
Firstly, if $\alpha^\star=0$ we get
\[
\lim_{n \to \infty} \exp\left\{\frac{\alpha_n c_n \log
n}{\left(1-c_n\frac{\log n}{n}\right)^{2}}\right\} =1
\]
and (\ref{eq:condition_to_zero_law}) readily follows.

Next, assume that $\alpha^\star>0$. Recalling the scaling
condition (\ref{eq:scalinglaw}), we write
\begin{eqnarray}
\lefteqn{q(\theta_n)+(1-q(\theta_n))\exp\left\{\frac{\alpha_n c_n
\log n}{\left(1-c_n\frac{\log n}{n}\right)^{2}}\right\}}
\\ \nonumber
&=& q(\theta_n)+(1-q(\theta_n))\alpha_n \log n
\frac{\exp\left\{\frac{\alpha_n c_n \log n}{\left(1-c_n\frac{\log
n}{n}\right)^{2}}\right\}}{\alpha_n \log n}
\\ \nonumber
&=& q(\theta_n) + c_n \log n^2 \cdot \frac{n^{-1+\frac{\alpha_n
c_n}{\left(1-c_n\frac{\log n}{n}\right)^{2}}}}{\alpha_n \log n}
\\ \nonumber
&\leq& q(\theta_n) + c_n \log n^2 \cdot \frac{n^{-1+\frac{
c_n}{\left(1-c_n\frac{\log n}{n}\right)^{2}}}}{\alpha_n \log n}
\end{eqnarray}
upon using the fact that $\alpha_n \leq 1$ in the last step. Under
the enforced assumptions, we clearly have
\[
\lim_{n \to \infty} \\\left(-1+\frac{ c_n}{\left(1-c_n\frac{\log
n}{n}\right)^{2}}\right)=-1+c<0
\]
and we find
\[
\lim_{n \to \infty} \left(c_n \log n^2 \cdot \frac{n^{-1+\frac{
c_n}{\left(1-c_n\frac{\log n}{n}\right)^{2}}}}{\alpha_n \log
n}\right)=0
\]
upon recalling the assumption that $\lim_{n \to \infty} \alpha_n
\log n = \alpha^\star>0$. The desired condition
(\ref{eq:condition_to_zero_law}) follows as we note that
$q(\theta_n)\leq 1$. \myendpf

\section{A proof of Theorem \ref{thm:OneLaw+Connectivity} (Outline)}
\label{sec:ProofConnectivityI}

Fix $n=2,3, \ldots $ and consider $\Theta = (K,P,\alpha)$ with
$\alpha$ in $(0,1)$ and positive integers $K,P$ such that $K \leq
P$. We define the events
\[
C_n(\Theta) := \left [ \mathbb{K \cap G}(n; \Theta)
\mbox{~is~connected} \right ]
\]
and
\[
I ( n ; \Theta) := \left [ \mathbb{K \cap G}(n;\Theta)
\mbox{~contains~no~isolated~nodes} \right ].
\]
If the random graph $\mathbb{K \cap G}(n; \Theta)$ is connected,
then it does not contain any isolated node, whence $C_n(\Theta)$
is a subset of $I(n;\Theta)$, and the conclusions
\begin{equation}
\bP{ C_n(\Theta) } \leq \bP{ I(n;\Theta) }
\label{eq:FromConnectivityToNodeIsolation1}
\end{equation}
and
\begin{equation}
\bP{ C_n(\Theta)^c } = \bP{ C_n(\Theta)^c \cap I (n;\theta) } +
\bP{ I(n;\Theta)^c } \label{eq:FromConnectivityToNodeIsolation2}
\end{equation}
obtain.

Taken together with Theorem \ref{thm:OneLaw+NodeIsolation}, the
relations (\ref{eq:FromConnectivityToNodeIsolation1}) and
(\ref{eq:FromConnectivityToNodeIsolation2}) pave the way to
proving Theorem \ref{thm:OneLaw+Connectivity}. Indeed, pick a
scaling $\Theta: \mathbb{N}_0 \rightarrow \mathbb{N}_0 \times
\mathbb{N}_0 \times (0,1) $ such that (\ref{eq:scalinglaw}) holds
for some $c>0$ and $\lim_{n \to \infty}\alpha_n \log n =
\alpha^\star$ exists. If $c < 1$, then $\lim_{n \rightarrow
\infty} \bP{ I(n;\Theta_n) } = 0$ by the zero-law for the absence
of isolated nodes, whence $\lim_{n \rightarrow \infty} \bP{
C_n(\Theta_n) } = 0$ with the help of
(\ref{eq:FromConnectivityToNodeIsolation1}). If $c> 1$, then
$\lim_{n \rightarrow \infty} \bP{ I(n;\Theta_n) } = 1$ by the
one-law for the absence of isolated nodes, and the desired
conclusion $\lim_{n \rightarrow \infty } \bP{ C_n(\Theta_n) } = 1$
(or equivalently, $\lim_{n \rightarrow \infty } \bP{
C_n(\Theta_n)^c } = 0$) will follow via
(\ref{eq:FromConnectivityToNodeIsolation2}) if we show the
following:

\begin{proposition}
{\sl For any scaling $\Theta: \mathbb{N}_0 \rightarrow
\mathbb{N}_0 \times \mathbb{N}_0 \times (0,1) $ such that
(\ref{eq:scalinglaw}) holds for some $c>1$,
 we have
\begin{equation}
\lim_{n \rightarrow \infty} \bP{ C_n(\Theta_n)^c \cap I
(n;\Theta_n) } = 0 . \label{eq:OneLawAfterReduction}
\end{equation}
as long as the condition
(\ref{eq:OneLaw+ConnectivityExtraCondition}) is satisfied. }
\label{prop:OneLawAfterReduction}
\end{proposition}

The basic idea in establishing Proposition
\ref{prop:OneLawAfterReduction} is to find a sufficiently tight
upper bound on the probability in (\ref{eq:OneLawAfterReduction})
and then to show that this bound goes to zero as $n$ becomes
large. This approach is similar to the one used for proving the
one-law for connectivity in ER graphs \cite[p. 164]{Bollobas}.

We begin by finding the needed upper bound: Fix $n=2,3, \ldots $
and consider $\Theta = (K,P,\alpha)$ with $\alpha$ in $(0,1)$ and
positive integers $K,P$ such that $K \leq P$. For the reasons that
will later become apparent we find it useful to introduce the
event $E_n(\boldsymbol{X}_n(\theta))$ in the following manner:
\begin{equation}
E_n(\boldsymbol{X}_n(\theta))= \bigcup_{S \subseteq \mathcal{N}: ~
|S| \geq 1} ~ \left[\left|\cup_{i \in S}
K_i(\theta)\right|~\leq~{X}_{n,|S|}(\theta)\right]
\label{eq:E_n_defn}
\end{equation}
where
$\boldsymbol{X}_n(\theta)=[{X}_{n,1}(\theta)~~{X}_{n,2}(\theta)~~
\cdots~~ {X}_{n,n}(\theta)]$ is an $n$-dimensional integer valued
array. Let
\[
r_n (\theta) := \min \left ( r(\theta), \left \lfloor \frac{n}{2}
\right \rfloor \right ) \quad {\rm with} \quad r(\theta) := \left
\lfloor \frac{P}{K} \right \rfloor.
\]
In due course, we always set
\begin{eqnarray}
X_{n,i}(\theta)= \left \{
\begin{array}{ll}
\lfloor \lambda K i\rfloor & ~ \mbox{$i=1,2,\ldots, r_n(\theta)$} \\
 & \\
\lfloor \mu P \rfloor &~ \mbox{$i=r_n(\theta)+1, \ldots, n$}
\end{array}
\right . \label{eq:X_S_theta}
\end{eqnarray}
for some $\lambda, \mu$ in $(0,\frac{1}{2})$ that will be
specified later. For convention, we also take $X_{n,0}=0$.

By a crude bounding argument we now get
\begin{eqnarray}
\lefteqn{\bP{ C_n(\Theta)^c \cap I(n;\Theta) }} &&
\\ \nonumber
 &\leq&
\bP{E_n(\boldsymbol{X}_n(\theta))} + \bP{ C_n(\Theta)^c \cap
I(n;\Theta) \cap E_n(\boldsymbol{X}_n(\theta))^c }
\end{eqnarray}
Hence, a proof of Proposition \ref{prop:OneLawAfterReduction}
consists of establishing the following two results.

\begin{proposition}
{\sl Consider a scaling $\Theta: \mathbb{N}_0 \rightarrow
\mathbb{N}_0 \times \mathbb{N}_0 \times (0,1) $ such that
(\ref{eq:scalinglaw}) holds for some $c>1$, and
(\ref{eq:OneLaw+ConnectivityExtraCondition}) is satisfied for some
$\sigma>0$. We have
\begin{equation}
\lim_{n \rightarrow \infty} \bP{E_n(\boldsymbol{X}_n(\theta_n))} =
0. \label{eq:OneLawAfterReductionPart1}
\end{equation}
where $\boldsymbol{X}_n(\theta_n)=[X_{n,1}(\theta_n) ~ \cdots ~
X_{n,n}(\theta_n)]$ is as specified in (\ref{eq:X_S_theta}) with
$\lambda $ in $(0, \frac{1}{2})$ is selected small enough to
ensure
\begin{equation}
\max \left ( 2 \lambda \sigma , \lambda \left( \frac{e^2}{\sigma}
\right) ^{\frac{ \lambda }{ 1 - 2 \lambda } } \right ) < 1,
\label{eq:ConditionOnLambda}
\end{equation}
and $\mu$ in $(0, \frac{1}{2})$ is selected so that
\begin{equation}
\max \left ( 2 \left ( \sqrt{\mu} \left ( \frac{e}{ \mu } \right
)^{\mu} \right )^\sigma, \sqrt{\mu} \left ( \frac{e}{ \mu }
\right)^{\mu} \right ) < 1 . \label{eq:ConditionOnMU+1}
\end{equation}
 }
\label{prop:OneLawAfterReductionPart1}
\end{proposition}
A proof of Proposition \ref{prop:OneLawAfterReductionPart1} can be
found in Section \ref{sec:OneLawAfterReductionPart1}. Note that
for any $\sigma
>0$, $\lim_{\lambda \downarrow 0} \lambda \left( \frac{e^2}{\sigma}
\right) ^{\frac{ \lambda }{ 1 - 2 \lambda } } =0 $ so that the
condition (\ref{eq:ConditionOnLambda}) can always be met by
suitably selecting $\lambda > 0$ small enough. Also, we have
$\lim_{\mu \downarrow 0} \left ( \frac{e}{ \mu } \right)^{\mu}
=1$, whence $\lim_{\mu \downarrow 0} \sqrt{\mu} \left ( \frac{e}{
\mu } \right)^{\mu} = 0$, and (\ref{eq:ConditionOnMU+1}) can be
made to hold for any $\sigma>0$ by taking $\mu > 0$ sufficiently
small.

\begin{proposition}
{\sl Consider a scaling $\Theta: \mathbb{N}_0 \rightarrow
\mathbb{N}_0 \times \mathbb{N}_0 \times (0,1) $ such that
(\ref{eq:scalinglaw}) holds for some $c>1$, and
(\ref{eq:OneLaw+ConnectivityExtraCondition}) is satisfied for some
$\sigma>0$. We have
\begin{equation}
\lim_{n \rightarrow \infty} \bP{ C_n(\Theta_n)^c \cap I
(n;\Theta_n) \cap E_n(\boldsymbol{X}_n(\theta_n))^ c } = 0 .
\label{eq:OneLawAfterReductionPart2}
\end{equation}
where $\boldsymbol{X}_n(\theta_n)=[X_{n,1}(\theta_n) ~ \cdots ~
X_{n,n}(\theta_n)]$ is as specified in (\ref{eq:X_S_theta}) with
$\mu$ in $(0, \frac{1}{2})$ is selected small enough to ensure
(\ref{eq:ConditionOnMU+1}) and $\lambda \in (0,\frac{1}{2})$ is
selected such that it satisfies (\ref{eq:ConditionOnLambda}).
\label{prop:OneLawAfterReductionPart2} }
\end{proposition}

A proof of Proposition \ref{prop:OneLawAfterReductionPart2} is
given in Section \ref{sec:OneLawAfterReductionPart2}.

\section{A proof of Proposition \ref{prop:OneLawAfterReductionPart1}}
\label{sec:OneLawAfterReductionPart1}

The arguments that will lead to
(\ref{eq:OneLawAfterReductionPart1}) are taken mostly from
\cite{YaganThesis}. First observe by a standard union bound that
\begin{eqnarray}\nonumber
\lefteqn{\bP{ E_n(\boldsymbol{X}_n(\theta)) }} &&
\\\nonumber
 &\leq & \sum_{ S
\subseteq \mathcal{N}: 1 \leq |S| \leq n } \bP{ \left|\cup_{i \in
S} K_i(\theta)\right|~\leq~{X}_{n,|S|}(\theta) } \nonumber \\
\nonumber &=& \sum_{r=1}^{ n } \left ( \sum_{S \in
\mathcal{N}_{n,r} } \bP{ \left|\cup_{i \in S}
K_i(\theta)\right|~\leq~{X}_{n,r}(\theta) } \right )
\end{eqnarray}
where $\mathcal{N}_{n,r} $ denotes the collection of all subsets
of $\{ 1, \ldots , n \}$ with exactly $r$ elements. By using
exchangeability and the fact that $|\mathcal{N}_{n,r} | = {n
\choose r}$, we get
\begin{eqnarray}\nonumber
\bP{ E_n(\boldsymbol{X}_n(\theta)) } & \leq& \sum_{r=1}^{ n } {n
\choose r}  \bP{ U_r(\theta)~\leq~{X}_{n,r}(\theta) }
\\ \label{eq:E_n_big_expression}
&=& \sum_{r=1}^{ r_n(\theta) } {n \choose r}  \bP{
U_r(\theta)~\leq~\lfloor \lambda r K\rfloor }
\\ \nonumber
&&~ + \sum_{r=r_n(\theta)+1}^{ n } {n \choose r} \bP{
U_r(\theta)~\leq~\lfloor \mu P \rfloor }
\end{eqnarray}
where $U_r(\theta)$ matches the definition given in
\cite{YaganMakowskiConnectivity}, i.e.,
\[
U_r(\theta) = |\cup_{i=1}^{r}K_i(\theta)|.
\]

Now, consider a scaling $\Theta: \mathbb{N}_0 \rightarrow
\mathbb{N}_0 \times \mathbb{N}_0 \times (0,1) $ such that
(\ref{eq:scalinglaw}) holds for some $c>1$ and assume that the
condition (\ref{eq:OneLaw+ConnectivityExtraCondition}) is
satisfied for some $\sigma>0$. Replace $\theta$ by $\theta_n$ in
(\ref{eq:E_n_big_expression}) with respect to this scaling. It was
shown in \cite[Proposition 7.4.14]{YaganThesis} that for any
scaling $\theta: \mathbb{N}_0 \rightarrow \mathbb{N}_0$ such that
(\ref{eq:OneLaw+ConnectivityExtraCondition}) holds for some
$\sigma>0$, we have
\begin{eqnarray}
\lim_{n \to \infty} \sum_{r= r_n(\theta_n)+1}^{n} {n \choose r} ~
\bP{ U_r(\theta_n) \leq \lfloor \mu P_n \rfloor} = 0
\label{eq:E_n_part2}
\end{eqnarray}
whenever $\mu$ in $(0,\frac{1}{2})$ is selected so that
(\ref{eq:ConditionOnMU+1}) holds. Hence, the desired conclusion
(\ref{eq:OneLawAfterReductionPart1}) will follow if we show that
\begin{equation}
\lim_{n \to \infty} \sum_{r= 1}^{r_n(\theta_n)} {n \choose r} ~
\bP{ U_r(\theta_n) \leq \lfloor \lambda r K_n \rfloor} = 0
\label{eq:E_n_part1}
\end{equation}
under the condition (\ref{eq:ConditionOnLambda}). Under the
enforced assumptions, one can easily deduce from \cite[Proposition
7.4.13]{YaganThesis} that for any $\lambda $ in $(0, \frac{1}{2})$
small enough to ensure (\ref{eq:ConditionOnLambda}) we have
(\ref{eq:E_n_part1}) and this establishes Proposition
\ref{prop:OneLawAfterReductionPart1}. \myendpf

\section{A proof of Proposition \ref{prop:OneLawAfterReductionPart2}}
\label{sec:OneLawAfterReductionPart2}

Fix $n=2,3, \ldots $ and consider $\Theta = (K,P,\alpha)$ with
$\alpha$ in $(0,1)$ and positive integers $K,P$ such that $K \leq
P$. For any non-empty subset $S$ of nodes, i.e., $S \subseteq \{1,
\ldots , n \}$, we define the graph $\mathbb{K \cap G} (n;\Theta)
(S)$ (with vertex set $S$) as the subgraph of $\mathbb{K \cap G}
(n;\Theta)$ restricted to the nodes in $S$. We also say that $S$
is {\em isolated} in $\mathbb{K \cap G} (n;\Theta)$ if there are
no edges (in $\mathbb{K \cap G}(n;\Theta)$) between the nodes in
$S$ and the nodes in the complement $S^c = \{ 1, \ldots , n \} -
S$. This is characterized by
\[
K_{i}(\theta) \cap K_{j}(\theta)
 = \emptyset \:\: \vee \:\: B_{ij}(\alpha) = 0 ,
\quad i \in S , \ j \in S^c  .
\]

With each non-empty subset $S$ of nodes, we associate several
events of interest: Let $C_n (\Theta ; S)$ denote the event that
the subgraph $\mathbb{K \cap G} (n;\Theta) (S)$ is itself
connected. The event $C_n (\Theta ; S)$ is completely determined
by the rvs $\{ K_i(\theta), \ i \in S \}$ and $\{ B_{ij}(\alpha),
\ i,j \in S \}$. We also introduce the event $B_n (\Theta ; S)$ to
capture the fact that $S$ is isolated in $\mathbb{K \cap G}
(n;\Theta)$, i.e.,
\begin{eqnarray}
\lefteqn{ B_n (\Theta ; S) } & &
\nonumber \\
&:= & \left [ K_{i}(\theta) \cap K_{j}(\theta)
 = \emptyset \:\: \vee
\:\: B_{ij}(p) = 0 , \quad i \in S , \ j \in S^c \right ] .
\nonumber
\end{eqnarray}
Finally, we set
\[
A_n (\Theta ; S) := C_n (\Theta ; S) \cap B_n (\Theta ; S) .
\]

The starting point of the discussion is the following basic
observation: If $\mathbb{K \cap G} (n;\Theta)$ is {\em not}
connected and yet has {\em no} isolated nodes, then there must
exist a subset $S$ of nodes with $|S| \geq 2$ such that $\mathbb{K
\cap G}(n;\Theta) (S)$ is connected while $S$ is isolated in
$\mathbb{K \cap G} (n;\Theta)$. This is captured by the inclusion
\begin{equation}
C_n(\Theta)^c \cap I(n;\Theta)\ \subseteq  \bigcup_{S \subseteq
\mathcal{N}: ~ |S| \geq 2} ~ A_n (\Theta ; S) \label{eq:BasicIdea}
\end{equation}
A moment of reflection should convince the reader that this union
need only be taken over all subsets $S$ of $\{1, \ldots , n \}$
with $2 \leq |S| \leq \lfloor \frac{n}{2} \rfloor $.

By a standard union bound argument, we immediately get
\begin{eqnarray}\nonumber
\lefteqn{\bP{ C_n(\Theta)^c \cap I(n;\Theta) \cap
E_n(\boldsymbol{X}_n(\theta))^c }}
\\ \nonumber
 &\leq & \sum_{ S \subseteq
\mathcal{N}: 2 \leq |S| \leq \lfloor \frac{n}{2} \rfloor } \bP{
A_n (\Theta ; S) \cap E_n(\boldsymbol{X}_n(\theta))^c }
\nonumber \\
&=& \sum_{r=2}^{ \lfloor \frac{n} {2} \rfloor } \left ( \sum_{S
\in \mathcal{N}_{n,r} } \bP{ A_n (\Theta ; S) \cap
E_n(\boldsymbol{X}_n(\theta))^c} \right )
\label{eq:BasicIdea+UnionBound}
\end{eqnarray}
where $\mathcal{N}_{n,r} $ denotes the collection of all subsets
of $\{ 1, \ldots , n \}$ with exactly $r$ elements.

For each $r=1, \ldots , n$, we simplify the notation by writing
$A_{n,r} (\Theta) := A_n (\Theta ; \{ 1, \ldots , r \} )$,
$B_{n,r} (\Theta) := B_n (\Theta ; \{ 1, \ldots , r \} )$ and
$C_{n,r}(\Theta) := C_n (\Theta ; \{ 1, \ldots , r \} )$. With a
slight abuse of notation, we use $C_n(\Theta)$ for $r=n$ as
defined before. Under the enforced assumptions, exchangeability
yields
\[
\bP{ A_n (\Theta ; S) } = \bP{ A_{n,r} (\Theta) }, \quad S \in
\mathcal{N}_{n,r}
\]
and the expression
\begin{eqnarray} \nonumber
\lefteqn{\sum_{S \in \mathcal{N}_{n,r} } \bP{ A_n (\Theta ; S)
\cap E_n(\boldsymbol{X}_n(\theta))^c }} &&
\\
&=& {n \choose r} ~ \bP{ A_{n,r}(\Theta ) \cap
E_n(\boldsymbol{X}_n(\theta))^c } \label{eq:ForEach=r}
\end{eqnarray}
follows since $|\mathcal{N}_{n,r} | = {n \choose r}$. Substituting
into (\ref{eq:BasicIdea+UnionBound}) we obtain the key bound
\begin{eqnarray}\nonumber
\lefteqn{\bP{ C_n(\Theta)^c \cap I(n;\Theta) \cap
E_n(\boldsymbol{X}_n(\theta))^c }} &&
\\
 &\leq& \sum_{r=2}^{ \lfloor
\frac{n}{2} \rfloor } {n \choose r} ~ \bP{ A_{n,r}(\Theta ) \cap
E_n(\boldsymbol{X}_n(\theta))^c} .
\label{eq:BasicIdea+UnionBound2}
\end{eqnarray}

Consider a scaling $\Theta: \mathbb{N}_0 \rightarrow \mathbb{N}_0
\times \mathbb{N}_0 \times (0,1)$ as in the statement of
Proposition \ref{prop:OneLawAfterReduction}. Substitute $\Theta$
by $\Theta_n$ by means of this scaling in the right hand side of
(\ref{eq:BasicIdea+UnionBound2}). The proof of Proposition
\ref{prop:OneLawAfterReduction} will be completed once we show
\begin{equation}
\lim_{n \rightarrow \infty} \sum_{r=2}^{ \lfloor \frac{n}{2}
\rfloor } {n \choose r} ~ \bP{ A_{n,r}(\Theta ) \cap
E_n(\boldsymbol{X}_n(\theta))^c} = 0. \label{eq:OneLawToShow}
\end{equation}
The means to do so are provided in the next section.

\section{Bounding the probabilities $\bP{A_{n,r}(\Theta)}$ \\
         ($r=1, \ldots , n$)}
\label{sec:BoundingProbabilities}

Consider $\alpha$ in $(0,1)$ and positive integers $K$ and $P$
such that $K \leq P$. Fix $n=2,3, \ldots $ and pick $r=1, \ldots ,
n-1$. First, observe the equivalence
\[
B_{n,r}(\Theta ) = \left [ \left ( \cup_{i \in v_{r,j}(\alpha)}
K_i(\theta) \right ) \cap K_j(\theta) = \emptyset, \ j=r+1, \ldots
n \right ]
\]
where $v_{r,j}(\alpha)$ is as defined in (\ref{eq:v}). Hence,
under the enforced assumptions on the rvs $K_1(\theta), \ldots ,
K_n(\theta) $, we readily obtain the expression

\begin{eqnarray}\nonumber
\lefteqn{\bP{ B_{n,r}(\Theta ) ~~\Big | ~~\begin{array}{c}
  K_i(\theta), \ i=1, \ldots , r \\ B_{ij}(\alpha), \ i=1,\ldots,r, j=r+1,\ldots,n
  \\
\end{array}  }
} &&
\\ \nonumber
&=& \prod_{j=r+1}^n \left ( {{P- |\cup_{i \in v_{r,j}(\alpha)}
K_i(\theta)|} \choose {K}} \over {{P \choose K}} \right )
\hspace{2cm}
\end{eqnarray}

It is clear that the distributional properties of the term
$|\cup_{i \in v_{r,j}(\alpha)} K_i(\theta)|$ will play an
important role in efficiently bounding $\bP{B_{n,r}(\Theta)}$.
Note that it is always the case that
\begin{equation}
|\cup_{i \in v_{r,j}(\alpha)} K_i(\theta)| \geq K
\1{|v_{r,j}(\alpha)|>0}. \label{eq:U_r_Trivial_Bound}
\end{equation}
Also, on the event $E_n(\boldsymbol{X}_n(\theta))^c$, we have
\begin{equation}
|\cup_{i \in v_{r,j}(\alpha)} K_i(\theta)| \geq \left(
X_{n,|v_{r,j}(\alpha)|}(\theta)+1\right) \cdot
\1{|v_{r,j}(\alpha)|>0} \label{eq:U_r_Difficult_Bould}
\end{equation}
for each $j=r+1, \ldots, n$. Conditioning on the rvs $K_1(\theta),
\ldots , K_r(\theta) $ and $\{ B_{ij}(\alpha), \ i,j=1,\ldots,r\}$
(which determine the event $C_{r}(\Theta )$), we conclude via
(\ref{eq:U_r_Trivial_Bound})-(\ref{eq:U_r_Difficult_Bould}) that
\begin{eqnarray}\nonumber
\lefteqn{\bP{ A_{n,r}(\Theta ) \cap
E_n(\boldsymbol{X}_n(\theta))^c}
 }  &&
\\ \nonumber
 &=& \bP{ C_{r}(\Theta ) \cap B_{n,r}(\Theta) \cap
E_n(\boldsymbol{X}_n(\theta))^c }
 \\ \nonumber
&\leq& \bE{ \begin{array}{l}
 \1{ C_{r}(\Theta ) } \times \\
 \times \prod_{j=r+1}^n {
{{P- \max \{ K, X_{n,|v_{r,j}(\alpha)|}(\theta)+1\} \cdot
\1{|v_{r,j}(\alpha)|>0} } \choose {K} }\over{{P \choose K}}}
\end{array} } \label{eq:ComputePA_{n,r}1}
\end{eqnarray}

Observe that the event $C_{r}(\Theta )$ is independent from the
set-valued random variables $v_{r,j}(\alpha)$ for each $j=r+1,
\ldots, n$. Also, as noted before
$\{|v_{r,j}(\alpha)|\}_{j=r+1}^{n}$ are i.i.d.. Invoking these we
obtain

\begin{eqnarray}\nonumber
\lefteqn{\bP{ A_{n,r}(\Theta ) \cap
E_n(\boldsymbol{X}_n(\theta))^c}
 }  &&
\\ \nonumber
&\leq&
  \bE {
{ {P- \max \{ K, X_{n,|v_{r}(\alpha)|}(\theta)+1\}
 \1{|v_r(\alpha)|>0} } \choose
K }\over{{P \choose K}}}^{n-r}
\\
&&~ \times \bP{C_{r}(\Theta )}\label{eq:ComputePA_{n,r}}
\end{eqnarray}
with $v_r(\alpha)$ denoting a generic random variable identically
distributed with $v_{r,j}(\alpha), \; j=r+1, \ldots, n$, i.e.,
\begin{equation}
v_r(\alpha) =_{\mbox{st}} \mbox{Bin}(r,\alpha).
\label{eq:v_r_alpha}
\end{equation}

We now compute the expectation appearing at
(\ref{eq:ComputePA_{n,r}}) by using the definition
(\ref{eq:X_S_theta}).

\begin{lemma}
{\sl Consider $\theta=(K,P)$ with positive integers $K$ and $P$
such that $K \leq P$. With $\boldsymbol{X}_n(\theta)$ defined as
in (\ref{eq:X_S_theta}) for some $\lambda$ and $\mu$ in $(0,
\frac{1}{2})$, we have
\begin{eqnarray}\nonumber
 \lefteqn{\bE {
{ {P- \max \{ K, X_{n,|v_{r}(\alpha)|}(\theta)+1\}
 \1{|v_r(\alpha)|>0} } \choose
K }\over{{P \choose K}}}} &&
\\
 & & ~ \leq
e^{-\alpha(1-q(\theta))\lambda r}+ e^{-K\mu} \1{r>r_n(\theta)}
\label{eq:crucial_bound_expectation}
\end{eqnarray}
for each $r=1, \ldots, \lfloor \frac{n}{2} \rfloor$.}
\label{lem:bounding_expectation}
\end{lemma}

\proof Fix $r=1, \ldots, r_n(\theta)$ and  recall
(\ref{eq:prelimiary}). On that range, we have
\[
\left( X_{|v_r(\alpha)|}(\theta)+1\right) \cdot
\1{|v_r(\alpha)|>0} \geq \lceil \lambda |v_r(\alpha)| K \rceil.
\]
Thus, in view of (\ref{eq:prelimiary}), we get
\begin{eqnarray}
\lefteqn{\bE { { {P- \max \{ K, X_{n,|v_{r}(\alpha)|}(\theta)+1\}
 \1{|v_r(\alpha)|>0} } \choose
K }\over{{P \choose K}}}} && \nonumber \\
&\leq& \bE { {{P- \max \{ \lceil \lambda |v_r(\alpha)| K \rceil },
K \1{|v_r(\alpha)|>0} \} \choose K}\over{{P \choose K}} }
\label{eq:int_3}\\
&\leq& \bE{q(\theta)^{\lambda |v_r(\alpha)| \1{\lambda
|v_r(\alpha)| \geq 1}+ \1{ |v_r(\alpha)| > 0} \1{\lambda
|v_r(\alpha)| < 1}}}.
\nonumber\\
&\leq& \bE{q(\theta)^{\lambda |v_r(\alpha)| \1{\lambda
|v_r(\alpha)| \geq 1}+  \lambda |v_r(\alpha)|  \1{\lambda
|v_r(\alpha)| < 1}}}. \label{eq:int_1}
\\ \label{eq:int_2} &=&
\bE{q(\theta)^{\lambda |v_r(\alpha)| }}
\end{eqnarray}
where in (\ref{eq:int_1}) we used the facts that
\[
\1{ |v_r(\alpha)| > 0} \1{\lambda |v_r(\alpha)| < 1} \geq \lambda
|v_r(\alpha)|  \1{\lambda |v_r(\alpha)| < 1}
\]
and $q(\theta)<1$.

 Now, direct computation via (\ref{eq:v_r_alpha}) yields
\begin{eqnarray}\nonumber
\bE {q(\theta)^{\lambda |v_r(\alpha)|} }
 &=& \sum_{j=0}^{r} {r \choose j} \alpha^j (1-\alpha)^{r-j}
 q(\theta)^{\lambda j}
 \\ \nonumber
  &=&\left( 1 -\alpha \left(1-q(\theta)^\lambda
\right)\right)^r
\\ \label{eq:new1}
&\leq& \left( 1 -\alpha \lambda \left(1-q(\theta)\right)\right)^r
\\ \label{eq:prob_expectation_bounded}
&\leq& e^{-\alpha (1-q(\theta)) \lambda r},
\end{eqnarray}
upon using (\ref{eq:prelimiaryB}) in (\ref{eq:new1}).

On the range $r=r_n(\theta)+1, \ldots, \lfloor \frac{n}{2}
\rfloor$, we use
\[
{{ {P-L} \choose {K} } \over { {P} \choose {K} }} \leq
e^{-\frac{K}{P} \cdot L}
\]
that holds \cite[Lemma 5.4.1, pp. 79]{YaganThesis} for any
positive integer $L$. Thus, in view of (\ref{eq:int_2}), we have
\begin{eqnarray}\nonumber\lefteqn{\bE {
{ {P- \max \{ K, X_{n,|v_{r}(\alpha)|}(\theta)+1\}
 \1{|v_r(\alpha)|>0} } \choose
K }\over{{P \choose K}}}} &&
\\ \label{eq:big_expectation_1}
&\leq& \bE {q(\theta)^{\lambda |v_r(\alpha)| } \1{|v_r(\alpha)|
\leq r_n(\theta)}}
\\ \nonumber
 & & ~ + \bE{e^{-\frac{K}{P}\cdot \left(
\lfloor \mu P \rfloor +1\right)  } \1{|v_r(\alpha)|
> r_n(\theta)} }
\end{eqnarray}
upon using the fact that
\[
\max \{ K, X_{n,|v_{r}(\alpha)|}(\theta)+1\}
 \1{|v_r(\alpha)|>0} \geq \lfloor \mu P \rfloor +1
\]
whenever $|v_r(\alpha)|
> r_n(\theta)$.

In view of (\ref{eq:prob_expectation_bounded}) and
(\ref{eq:big_expectation_1}), we now obtain
\begin{eqnarray}
\lefteqn{\bE { { {P- \max \{ K, X_{n,|v_{r}(\alpha)|}(\theta)+1\}
 \1{|v_r(\alpha)|>0} } \choose
K }\over{{P \choose K}} }} &&
 \nonumber \\
&\leq& e^{-\alpha (1-q(\theta)) \lambda r}
\\ \nonumber
& & ~ +  \sum_{j=r_n(\theta)+1}^{r} {r \choose j} \alpha^j
 (1-\alpha)^{r-j}e^{-\frac{K}{P} (\lfloor \mu P \rfloor + 1)  }
 \\ \nonumber
 &\leq& e^{-\alpha (1-q(\theta)) \lambda r} + \sum_{j=0}^{r} {r \choose j} \alpha^j
 (1-\alpha)^{r-j} e^{-\mu K}
 \\ \nonumber
 &\leq& e^{-\alpha (1-q(\theta)) \lambda r} + e^{-\mu K}
\end{eqnarray}
and the desired conclusion (\ref{eq:crucial_bound_expectation})
follows.
 \myendpf

We also find it useful to note the crude bound
\begin{eqnarray}\nonumber
\lefteqn{\bP{ A_{n,r}(\Theta ) \cap
E_n(\boldsymbol{X}_n(\theta))^c}
 }  &&
\\ \label{eq:big_expectation_2}
&\leq& \bP{C_{r}(\Theta )}  \bE
{q(\theta)^{\1{|v_r(\alpha)|>0}}}^{n-r}
\end{eqnarray}
immediate by direct inspection from (\ref{eq:ComputePA_{n,r}}) and
(\ref{eq:int_3}).

The next result shows that for each $r=2. \ldots , n$, the
probability of the event $C_{r} (\Theta)$ can be provided an upper
bound in terms of known quantities.
\begin{lemma}
{\sl For each $r=2, \ldots , n$, we have
\begin{equation}
\bP{ C_{r} (\Theta) } \leq r^{r-2} \left ( \alpha\left(1 -
q(\theta)\right) \right)^{r-1} . \label{eq:ProbabilityOfC}
\end{equation}
} \label{lem:ProbabilityOfC}
\end{lemma}

\myproof First some notation: For each $r=2, \ldots ,n$, let
$\mathbb{K}_r(n;\theta)$ and $\mathbb{G}_r(n;\alpha)$ define the
subgraphs $\mathbb{K}(S)$ and $\mathbb{G}(S)$, respectively, when
$S = \{1 , \ldots , r \}$. Similarly let
 $\mathbb{K}_r\cap\mathbb{G}_r(n;\Theta)$ stand for
the subgraph $\mathbb{K}\cap\mathbb{G}(S)$. Finally, let ${\cal
T}_{r}$ denote the collection of all spanning trees on the vertex
set $\{1, \ldots , r \}$. It was shown \cite[Lemma 7.4.5, pp.
124]{YaganThesis} that
\begin{equation}
\bP{ T \subset \mathbb{K}_r(n;\theta) } = \left ( 1 - q(\theta)
\right )^{r-1}, \quad T \in {\cal T}_{r}
\label{eq:ProbabilityOfTree}
\end{equation}
where the notation $T \subset \mathbb{K}_r(n;\theta)$ indicates
that the tree $T$ is a subgraph spanning $\mathbb{K}_r(n;\theta)$.
It is also well known \cite{Bollobas} that
\begin{equation}
\bP{ T \subset \mathbb{G}_r(n;\alpha) } = \alpha^{r-1}, \quad T
\in {\cal T}_{r} \label{eq:ProbabilityOfTree2}
\end{equation}
By independence we find
\begin{equation}
\bP{ T \subset \mathbb{K}_r\cap \mathbb{G}_r(n;\Theta) } = \left (
\alpha \left( 1 - q(\theta) \right) \right )^{r-1}, \quad T \in
{\cal T}_{r}, \label{eq:ProbabilityOfTree3}
\end{equation}
upon combining (\ref{eq:ProbabilityOfTree}) and
(\ref{eq:ProbabilityOfTree2}).

By Cayley's formula \cite{Martin} there are $r^{r-2}$ trees on $r$
vertices, i.e., $| {\cal T}_{r}| = r^{r-2}$, and
(\ref{eq:ProbabilityOfC}) follows (via
(\ref{eq:ProbabilityOfTree3})) upon making use of a union bound.
\myendpf

\section{Establishing (\ref{eq:OneLawToShow})}

It is now clear how to proceed: Consider an admissible scaling
$K,P: \mathbb{N}_0 \rightarrow \mathbb{N}_0$ and a scaling
$\alpha: \mathbb{N}_0 \rightarrow (0,1)$ as in the statement of
Proposition \ref{prop:OneLawAfterReduction}. 
For the time being, pick an integer $R \geq 2$ (to be specified in
Section \ref{sec:Last_Parts_2}), and on the range $n \geq
n^\star(R)$ consider the decomposition
\begin{eqnarray}
\lefteqn{\sum_{r=2}^{\lfloor \frac{n}{2} \rfloor} {n \choose r} ~
\bP{ A_{n,r} (\Theta_n) \cap E_n(\boldsymbol{X}_n(\theta_n))^c } }
&&
\\ \nonumber
&=& \hspace{-2mm}\sum_{r=2}^{ R } {n \choose r} ~ \bP{ A_{n,r}
(\Theta_n) \cap E_n(\boldsymbol{X}_n(\theta_n))^c }
\label{eq:AnotherDecomposition}\\
&&+\sum_{r=R+1}^{ \max\{R,r_n (\theta)\} } { n \choose r} ~ \bP{
A_{n,r} (\Theta_n) \cap E_n(\boldsymbol{X}_n(\theta_n))^c }
\nonumber \\
&&+\hspace{-2mm}\sum_{r=\max\{R,r_n (\theta)\} +1}^{\lfloor
\frac{n}{2} \rfloor} \hspace{-2mm} { n \choose r} ~ \bP{ A_{n,r}
(\Theta_n) \cap E_n(\boldsymbol{X}_n(\theta_n))^c } . \nonumber
\end{eqnarray}
Let $n$ go to infinity: The desired convergence
(\ref{eq:OneLawToShow}) will be established if we show
\begin{equation}
\lim_{n \rightarrow \infty} \sum_{r=2}^{ R } { n \choose r} ~ \bP{
A_{n,r} (\Theta_n) \cap E_n(\boldsymbol{X}_n(\theta_n))^c } = 0 ,
\label{eq:StillToShow0}
\end{equation}
\begin{equation}
\lim_{n \rightarrow \infty} \sum_{r=R+1}^{ \max\{R,r_n
(\theta_n)\} } { n \choose r} ~ \bP{ A_{n,r} (\Theta_n) \cap
E_n(\boldsymbol{X}_n(\theta_n))^c } = 0 \label{eq:StillToShow1}
\end{equation}
and
\begin{eqnarray}
\lefteqn{\lim_{n \rightarrow \infty} \sum_{ r=\max\{R,r_n
(\theta_n)\} +1}^{\lfloor \frac{n}{2} \rfloor} { n \choose r} ~
\bP{ A_{n,r} (\Theta_n) \cap E_n(\boldsymbol{X}_n(\theta_n))^c }}
&&
\nonumber\\
&=& 0 \hspace{7cm}. \label{eq:StillToShow2}
\end{eqnarray}

The next sections are devoted to proving the validity of
(\ref{eq:StillToShow0}), (\ref{eq:StillToShow1}) and
(\ref{eq:StillToShow2}) by repeated applications of the
inequalities (\ref{eq:ComputePA_{n,r}}) and Lemmas
\ref{lem:bounding_expectation}-\ref{lem:ProbabilityOfC}.
 Throughout, we also make repeated use of the
standard bounds
\begin{equation}
{n \choose r} \leq \left ( \frac{e n}{r} \right )^r
\label{eq:CombinatorialBound1}
\end{equation}
valid for all $r,n=1,2, \ldots $ with $r\leq n$.
Finally, we note by convexity that the inequality
\begin{equation}
(x + y )^p \leq 2^{p-1} (x^p + y ^p ), \quad
\begin{array}{c}
x,y \geq 0 \\
p \geq 1
\end{array}
\label{eq:ConvexityInequality}
\end{equation}
holds.

\section{Establishing (\ref{eq:StillToShow0})}
\label{sec:Last_Parts_1}

For any arbitrary integer $R\geq 2$, it is clear that
(\ref{eq:StillToShow0}) will follow upon showing
\[
\lim_{n \rightarrow \infty} { n \choose r} ~ \bP{ A_{n,r}
(\Theta_n) \cap E_n(\boldsymbol{X}_n(\theta_n))^c } = 0, \quad
r=2,3, \ldots
\]
Fix $r=2,3, \ldots$ and recall (\ref{eq:big_expectation_2}),
$(\ref{eq:ProbabilityOfC})$ together with
(\ref{eq:CombinatorialBound1}). We get
\begin{eqnarray}\nonumber
\lefteqn{{ n \choose r} ~ \bP{ A_{n,r} (\Theta) \cap
E_n(\boldsymbol{X}_n(\theta))^c }} &&
\\ \nonumber
&\leq& { n \choose r}  \bP{C_{r}(\Theta)}  \bE
{q(\theta)^{\1{|v_r(\alpha)|>0}}}^{n-r}
\\ \nonumber
&\leq& \left(\frac{e n}{r}\right)^{r} r^{r-2} \left (
\alpha\left(1 - q(\theta)\right) \right)^{r-1}  \bE
{q(\theta)^{\1{|v_r(\alpha)|>0}}}^{n-r}
\end{eqnarray}
while we find
\begin{eqnarray}\nonumber
 \bE
{q(\theta)^{\1{|v_r(\alpha)|>0}}} &=& (1-\alpha)^r +
\left(1-(1-\alpha)^r\right)q(\theta)
\\ \nonumber
&\leq& 1-\alpha(1-q(\theta))
\end{eqnarray}
upon using (\ref{eq:v_r_alpha}) and the fact that $q(\theta) \leq
1$. This yields
\begin{eqnarray}\nonumber
\lefteqn{{ n \choose r} ~ \bP{ A_{n,r} (\Theta) \cap
E_n(\boldsymbol{X}_n(\theta))^c }} &&
\\ \label{eq:part_1_before_scaling}
&\leq& (e n)^{r} \left ( \alpha\left(1 - q(\theta)\right)
\right)^{r-1} \left(1-\alpha(1-q(\theta))\right)^{n-r}
\end{eqnarray}

Now consider a scaling $\Theta: \mathbb{N}_0 \rightarrow
\mathbb{N}_0 \times \mathbb{N}_0 \times (0,1)$ such that
(\ref{eq:scalinglaw}) holds for some $c>1$ and replace $\Theta$ by
$\Theta_n$ in (\ref{eq:part_1_before_scaling}) according to this
scaling. We find
\begin{eqnarray}\nonumber
\lefteqn{{ n \choose r} ~ \bP{ A_{n,r} (\Theta_n) \cap
E_n(\boldsymbol{X}_n(\theta_n))^c }} &&
\\ \nonumber
&\leq& (e n)^{r} \left ( \alpha_n\left(1 - q(\theta_n)\right)
\right)^{r-1} \left(1-\alpha_n(1-q(\theta_n))\right)^{n-r}
\\ \nonumber
&=& (e n)^{r} \left ( c_n \frac{\log n}{n} \right)^{r-1}
\left(1-c_n \frac{\log n}{n}\right)^{n-r}
\\ \nonumber
&\leq& n \left(e c_n \log n \right)^{r} e^{-c_n \log n
\frac{n-r}{n}}
\\ \nonumber
&=& \left(e c_n \log n \right)^{r} n^{1 - c_n \frac{n-r}{n}}
\end{eqnarray}
with the sequence $c: \mathbb{N}_0 \rightarrow \mathbb{R}_+$
satisfying $\lim_{n \rightarrow \infty} c_n = c>1$. Now let $n$
grow large in this last inequality. We obtain
\[
\lim_{n \to \infty} \left(1- c_n \frac{n-r}{n}\right) = 1- c < 0
\]
and the desired conclusion (\ref{eq:StillToShow0}) follows for any
$r=2,3, \ldots $. \myendpf

\section{Establishing (\ref{eq:StillToShow1})}
\label{sec:Last_Parts_2}

Consider a scaling $\Theta: \mathbb{N}_0 \rightarrow \mathbb{N}_0
\times \mathbb{N}_0 \times (0,1)$ and positive scalars $\lambda,
\mu$ as in the statement of Proposition
\ref{prop:OneLawAfterReductionPart2}. Since $R$ can be taken to be
arbitrarily large by virtue of the previous section, the desired
relation (\ref{eq:StillToShow1}) follows immediately if $\lim \sup
r_n(\theta_n) < \infty$. Assume now that $\lim \sup r_n(\theta_n)
= \infty$ and on the range $r=R+1, \ldots, r_n(\theta_n)$, recall
(\ref{eq:ComputePA_{n,r}}), (\ref{eq:crucial_bound_expectation}),
 and
(\ref{eq:ProbabilityOfC}). We get
\begin{eqnarray}
\lefteqn{ { n \choose r} ~ \bP{ A_{n,r} (\Theta_n) \cap
E_n(\boldsymbol{X}_n(\theta_n))^c }} &&
\\ \nonumber
&\leq& \left(\frac{e n}{r}\right)^{r} r^{r-2}\left(\alpha_n
(1-q(\theta_n))\right)^{r-1} e^{-\alpha_n (1-q(\theta_n)) r
\lambda (n-r)}
\\ \nonumber
&=&\left(\frac{e n}{r}\right)^{r} r^{r-2}\left(c_n \frac{\log
n}{n}\right)^{r-1} e^{-c_n \frac{\log n}{n} r \lambda (n-r)}
\\ \nonumber
&\leq& n \left( e c_n \log n\right)^{r} e^{- c_n \log n \cdot r
\lambda \frac{n-r}{n} }.
\end{eqnarray}
Now, observe that on the range $r=R+1, \ldots, r_n(\theta_n)$, we
have $r\leq \lfloor \frac{n}{2}\rfloor$ so that $\frac{n-r}{n}
\geq \frac{1}{2}$. This yields
\begin{eqnarray}\nonumber
\lefteqn{ \sum_{r=R+1}^{r_n(\theta_n)}{ n \choose r} ~ \bP{
A_{n,r} (\Theta_n) \cap E_n(\boldsymbol{X}_n(\theta_n))^c }} &&
\\ \nonumber
&\leq&  \sum_{r=R+1}^{r_n(\theta_n)} n \left( e c_n \log n ~
e^{-\lambda \frac{c_n}{2} \log n } \right)^{r}
\\ \label{eq:infintite_series_last_part}
&\leq&  \sum_{r=R+1}^{\infty} n \left( e c_n \log n ~ e^{-\lambda
\frac{c_n}{2} \log n } \right)^{r}.
\end{eqnarray}
Observe that
\begin{equation}
\lim_{n \to \infty} e c_n \log n ~ e^{-\frac{c_n}{2}  \lambda \log
n } = 0 \label{eq:infintite_series_to_zero}
\end{equation}
so that the infinite series appearing at
(\ref{eq:infintite_series_last_part}) is summable. Indeed, for $n$
sufficiently large to ensure that $ e c_n \log n ~
e^{-\frac{c_n}{2}  \lambda \log n} < 1$, we find
\begin{eqnarray}\nonumber
\lefteqn{ \sum_{r=R+1}^{r_n(\theta_n)}{ n \choose r} ~ \bP{
A_{n,r} (\Theta_n) \cap E_n(\boldsymbol{X}_n(\theta_n))^c }} &&
\\ \nonumber
&\leq&  n \frac{\left( e c_n \log n ~ e^{-\frac{c_n}{2}  \lambda
\log n } \right)^{R+1}}{1-e c_n \log n ~ e^{-\frac{c_n}{2} \lambda
\log n }}
\\ \nonumber
&=&   \frac{\left( e c_n \log n \right)^{R+1} n^{1-\frac{c_n}{2}
\lambda (R+1)}}{1-e c_n \log n ~ e^{-\frac{c_n}{2}\lambda \log n,
}}
\end{eqnarray}
where the sequence $c: \mathbb{N}_0 \rightarrow \mathbb{R}_+$
satisfies $\lim_{n \rightarrow \infty} c_n = c$.

Now let $n$ go to infinity in this last expression. In view of
(\ref{eq:infintite_series_to_zero}), we get
(\ref{eq:StillToShow1}) whenever $R$ is selected large enough to
satisfy
\begin{equation}
\frac{c}{2} \lambda (R+1) > 1. \label{eq:condition_on_R}
\end{equation}
Note that we have $c>1$ and $\lambda>0$. Thus,
(\ref{eq:condition_on_R}) can always be satisfied by selecting
\begin{equation}
R \geq \frac{2}{\lambda} \label{eq:R}
\end{equation}
and (\ref{eq:StillToShow1}) is now established. \myendpf

\section{Establishing (\ref{eq:StillToShow2})}
\label{sec:Last_Parts_3}

Consider a scaling $\Theta: \mathbb{N}_0 \rightarrow \mathbb{N}_0
\times \mathbb{N}_0 \times (0,1)$ and positive scalars $\lambda,
\mu$ as in the statement of Proposition
\ref{prop:OneLawAfterReductionPart2}. On the range $r=\max \{R ,
r_n(\theta_n)\}+1, \ldots, \lfloor \frac{n}{2}\rfloor $, recall
(\ref{eq:ComputePA_{n,r}}), (\ref{eq:crucial_bound_expectation}),
and (\ref{eq:ProbabilityOfC}). We get

\begin{eqnarray}\nonumber
\lefteqn{ { n \choose r} ~ \bP{ A_{n,r} (\Theta_n) \cap
E_n(\boldsymbol{X}_n(\theta_n))^c }} &&
\\ \nonumber
&\leq& { n \choose r}  \bP{C_{r}(\Theta_n)} \left( e^{-\alpha_n
(1-q(\theta_n)) r  \lambda} + e^{-K_n\mu}\right) ^{n-r}
\\ \label{eq:last_step_3}
&\leq& { n \choose r}  \bP{C_{r}(\Theta_n)} \left( e^{-c_n
\frac{\log n}{n} r \lambda } + e^{-K_n\mu}\right) ^{\frac{n}{2}}
\end{eqnarray}
where the sequence $c: \mathbb{N}_0 \rightarrow \mathbb{R}_+$
satisfies $\lim_{n \rightarrow \infty} c_n = c>1$.

 We will establish
(\ref{eq:StillToShow2}) in two steps. First set
\[
\hat{r}_n = \left \lceil \frac{3}{\lambda} \frac{n}{\log n} \right
\rceil.
\]
Obviously, the range $r=\max\{R,r_n(\theta_n)\}+1, \ldots, \lfloor
\frac{n}{2} \rfloor $ is intersecting the range $r=\hat{r}_n,
\ldots, \lfloor \frac{n}{2} \rfloor $. For the latter range, we
invoke (\ref{eq:last_step_3}) to get
\begin{eqnarray}\nonumber
\lefteqn{ \sum_{r=\hat{r}_n}^{\lfloor \frac{n}{2}\rfloor}{ n
\choose r} ~ \bP{ A_{n,r} (\Theta_n) \cap
E_n(\boldsymbol{X}_n(\theta_n))^c }} &&
\\ \nonumber
&\leq& \sum_{r=\hat{r}_n}^{\lfloor \frac{n}{2}\rfloor}  { n
\choose r}  \left( e^{-c_n \frac{\log n}{n} r \lambda } +
e^{-K_n\mu}\right) ^{\frac{n}{2}}
\\ \nonumber
&\leq& \sum_{r=\hat{r}_n}^{\lfloor \frac{n}{2}\rfloor}  { n
\choose r}  \left( e^{-3} + e^{-K_n\mu}\right) ^{\frac{n}{2}}
\end{eqnarray}
for $n$ sufficiently large since $\lim_{n \to \infty}c_n = c > 1$.
Using the binomial formula
\begin{equation}
\sum_{r= \hat{r}_n }^{\lfloor \frac{n}{2} \rfloor} {n \choose r}
\leq 2^n, \label{eq:Bin}
\end{equation}
this yields
\begin{eqnarray}\nonumber
\lefteqn{ \sum_{r=\hat{r}_n}^{\lfloor \frac{n}{2}\rfloor}{ n
\choose r} ~ \bP{ A_{n,r} (\Theta_n) \cap
E_n(\boldsymbol{X}_n(\theta_n))^c }} &&
\\ \nonumber
&\leq& 2^n \left(  e^{-3} + e^{-K_n\mu}\right) ^{\frac{n}{2}}
\\ \nonumber
&\leq& (2\sqrt{2})^n \left( e^{-\frac{3}{2}n} +
e^{-\frac{K_n\mu}{2} n} \right)
\end{eqnarray}
upon also invoking (\ref{eq:ConvexityInequality}). Now, let $n$ go
to infinity and recall from (\ref{eq:usefulcons3}) that $\lim_{n
\to \infty} K_n = \infty$. We immediately get
\begin{eqnarray}\label{eq:last_step_3b}
\lim_{n \to \infty} \sum_{r=\hat{r}_n}^{\lfloor
\frac{n}{2}\rfloor}{ n \choose r} ~ \bP{ A_{n,r} (\Theta_n) \cap
E_n(\boldsymbol{X}_n(\theta_n))^c } =0
\end{eqnarray}
since $ 2 \sqrt{2} \cdot e^{-\frac{3}{2}} < 1$.

If $\hat{r}_n \leq r_n(\theta_n)+1$ for all $n$ sufficiently
large, then the desired condition (\ref{eq:StillToShow2}) is
automatically satisfied via (\ref{eq:last_step_3b}). On the other
hand, if $ r_n(\theta_n)+1 < \hat{r}_n $, we should still consider
the range $r=\max\{R,r_n(\theta_n)\}+1, \ldots, \hat{r}_n$. But,
on that range we have
\begin{eqnarray}
\lefteqn{e^{-c_n \frac{\log n}{n} r \lambda } + e^{-\mu K_n}} &&
\nonumber \\
&=& e^{-c_n \frac{\log n}{n} r \lambda } \left(1+ e^{-\mu K_n +
c_n \frac{\log n}{n} r \lambda}\right)
\nonumber \\
&\leq& \exp\left\{-c_n \frac{\log n}{n} r \lambda + e^{-\mu
K_n+c_n \frac{\log n}{n} r \lambda}\right\}
\nonumber \\
&=& \exp\left\{-c_n \frac{\log n}{n} r \lambda \left(1-
\frac{e^{-\mu K_n+c_n \frac{\log n}{n} r \lambda}}{c_n \frac{\log
n}{n} r \lambda}\right)\right\} \nonumber \\
&\leq& \exp\left\{-c_n \frac{\log n}{n} r \lambda \left(1-
\frac{e^{-\mu K_n+3 c_n}}{c_n \frac{\log n}{n} r
\lambda}\right)\right\}
 \label{eq:last}
\end{eqnarray}
while it also holds that
\begin{eqnarray}
\frac{e^{-\mu K_n}}{c_n \frac{\log n}{n} r \lambda} \leq
\frac{e^{-\mu K_n}}{c_n \frac{\log n}{n} \frac{P_n}{K_n} \lambda}
\leq \frac{K_n e^{-\mu K_n}}{c_n \sigma \lambda}.
\end{eqnarray}
Invoking the consequence (\ref{eq:usefulcons3}) yields
\[
\lim_{n \to \infty} K_n e^{-\mu K_n} = 0,
\]
whence we get
\[
\lim_{n \to \infty} \frac{e^{-\mu K_n + 3 c_n}}{c_n \frac{\log
n}{n} r \lambda} = 0.
\]

It is now immediate via (\ref{eq:last}) that for any given
$\epsilon > 0$, there exists a finite integer $n^\star$ such that
if $n \geq n^\star$, we have
\[
e^{-c_n \frac{\log n}{n} r \lambda}  + e^{-\mu K_n} \leq e^{-c_n
\frac{\log n}{n} r \lambda (1-\epsilon)}.
\]
Thus, on the range $n=n^\star+1, \ldots$, we use
(\ref{eq:last_step_3}) to get
\begin{eqnarray}\nonumber
\lefteqn{ \sum_{\max\{R.r_n(\theta_n)\}+1}^{\hat{r}_n} { n \choose
r} ~ \bP{ A_{n,r} (\Theta_n) \cap
E_n(\boldsymbol{X}_n(\theta_n))^c }} &&
\\ \nonumber
&\leq& \sum_{\max\{R,r_n(\theta_n)\}+1}^{\hat{r}_n} { n \choose r}
~ \bP{ C_{r} (\Theta_n)} e^{-c_n \frac{\log n}{n} r \lambda
(1-\epsilon)\frac{n}{2}}
\end{eqnarray}

Arguments leading to (\ref{eq:infintite_series_last_part}) gives
\begin{eqnarray}\nonumber
\lefteqn{ \sum_{\max\{R,r_n(\theta_n)\}+1}^{\hat{r}_n}  { n
\choose r} ~ \bP{ A_{n,r} (\Theta_n) \cap
E_n(\boldsymbol{X}_n(\theta_n))^c }} &&
\\ \nonumber
&\leq&  \sum_{r=\max\{R,r_n(\theta_n)\}+1}^{\infty} n \left( e c_n
\log n ~ e^{-c_n\frac{(1-\epsilon)}{2}  \lambda \log n }
\right)^{r}
\end{eqnarray}
and via similar arguments it is easy to see that
\[
\lim_{n \to \infty} \sum_{r=\max\{R,r_n(\theta_n)\}+1}^{\infty} n
\left( e c_n \log n ~ e^{-\frac{c_n(1-\epsilon)}{2} \lambda \log n
} \right)^{r} =0
\]
as long as
\[
\liminf_{n \to \infty} \frac{(1-\epsilon)}{2} c \lambda
\max\{R,r_n(\theta_n)\} > 1
\]
The above relation can be guaranteed by choosing $R$ such that
\[
R \geq \frac{2}{\lambda}
\]
as in (\ref{eq:R}). The desired conclusion (\ref{eq:StillToShow2})
is now established. \myendpf

\section*{Acknowledgment}
This work was supported by NSF Grant CCF-07290.

\bibliographystyle{IEEE}


\end{document}